
\def\Im{\,{\rm Im}\,}
\def\Re{\,{\rm Re}\,}
\def\pmb#1{\setbox0=\hbox{$#1$}
\kern-.025em\copy0\kern-\wd0
\kern.05em\copy0\kern-\wd0
\kern-.025em\raise.0433em\box0 }
\def\svec#1{\skew{-2}\vec#1}
\magnification=1200
\hoffset=-.1in
\voffset=-.2in

\vsize=7.5in
\hsize=5.6in
\tolerance 10000
\input eqno

\baselineskip 12pt plus 1pt minus 1pt
\pageno=0
\centerline{\bf The $\pmb{\Lambda N - \Sigma N}$ INTERACTION WITH ISOBAR}
\smallskip
\centerline{{\bf COUPLING AND SIX-QUARK RESONANCES}
\footnote{*}{This work is supported in part by funds
provided by the U. S. Department of Energy (D.O.E.) under contract
\#DE-AC02-76ER03069.}}
\vskip 24pt
\centerline{W. R. Greenberg}
\vskip 12pt
\centerline{\it Department of Physics, FM--15}
\centerline{\it University of Washington}
\centerline{\it Seattle, Washington\ \ 98195\ \ \ U.S.A.}
\vskip 12pt
\centerline{and}
\vskip 12pt
\centerline{E. L. Lomon}
\vskip 12pt
\centerline{\it Center for Theoretical Physics}
\centerline{\it Laboratory for Nuclear Science}
\centerline{\it and Department of Physics}
\centerline{\it Massachusetts Institute of Technology}
\centerline{\it Cambridge, Massachusetts\ \ 02139\ \ \ U.S.A.}
\vskip 1.5in
\centerline{Submitted to: {\it Physical Review D\/}}
\vfill
\centerline{ Typeset in $\TeX$ by Roger L. Gilson}
\vskip -12pt
\noindent CTP\#2065\hfill Revised December 1992
\eject
\baselineskip 24pt plus 2pt minus 2pt
\centerline{\bf ABSTRACT}
\medskip
The long-range $\Lambda N-\Sigma N$ interaction is modeled by a
configuration space meson-exchange potential matrix coupling to channels with
$\Delta$ and $\Sigma(1385)$ isobars.  An inner boundary condition, based on
$R$-matrix theory, replaces form factors for short-range effects and includes
the effects of free quark configurations.  An excellent fit is obtained to the
available data, with only the energy-independent boundary conditions as free
parameters.  The effect of isobar thresholds is shown to be substantial in
several partial waves and is crucial to the understanding of the higher energy
$\Lambda N$ elastic scattering data.
The positions and widths of $\left[q\left(
1s_{1/2}\right)\right]^5$ $s\left(1s_{1/2}\right)$ quark exotics are
predicted.
\vfill
\eject
\noindent{\bf I.\quad INTRODUCTION}
\medskip
\nobreak
The confinement property of QCD leads one to expect that the long-range
interaction between hadrons is mediated by hadron exchange.  Experience shows
that this is certainly so beyond $\sim 1.5$~fm inter-hadron separation, and
lattice QCD calculations indicate such behavior.  At short enough range the
asymptotic freedom property of QCD implies that simple quark configurations
will dominate over hadronic clusters of quarks, so that the hadron exchange
interaction representation will be inadequate or at least much more
complicated.  A conventional minimal allowance for the breakdown of the
hadronic description is the use of form factors with ``cut-off'' parameters
for each vertex.  This approach, using meson exchange potentials in momentum
space, has been used for several two-particle systems by the KFA--J\"ulich
group, including a successful fit to the $\Lambda N - \Sigma N$ system.$^1$
In our view this is not the best description of the transition from the
hadronic to the quark-gluon phase, and in particular does not allow for the
effect of levels of the six-quark configurations.

Hybrid approaches in which hadronic degrees of freedom are replaced by quark
degrees of freedom at small distances have been used with some success for the
nucleon-nucleon system.  In particular, the $R$-matrix method has been
successful in producing quantitative fits to a broad range of two-nucleon
data$^{2,\,3}$ and has also predicted six-quark ``exotics''$^{4-6}$ for which
there is some experimental evidence.$^{5-7}$  $R$-matrix theory provides
powerful relations between states of the inner region and the $S$-matrix of
the outer region, yielding quantitative results with relatively simple
calculations.  This allows the inclusion of the most important physics
ingredients.  $R$-matrix theory is formulated in configuration space, which is
also appropriate to the expected rapid transition as a function of distance
between the two limits of QCD.$^8$

This method is extended here to the $\Lambda N - \Sigma N$ system.  We also
recognize from the experience with the $NN$ system, that coupling to channels
with isobar components has an important influence well below their thresholds.
 Furthermore, the exotic excitations are at higher energies than the lowest
isobar thresholds, so that the energy dependence due to the threshold branch
points must be treated accurately.  The only way to do so is with an explicit
coupled channel calculation, as we do here in coupling to channels with the
$\Delta(1232)$ or the $\Sigma(1385)$.

Section~II presents the meson exchange potential in configuration space and
details the matrix components.  The matrix
elements of the spin and isospin operators are given.  In Section~III we
briefly review the $R$-matrix theory basis of the boundary condition at the
core surface, and the form of the boundary condition required by QCD.  The
coupled channel equations for the partial wave amplitudes are specified in
Section~IV, together with the expressions for the amplitudes and observables.

The results and the fitted core parameters are presented in Section V.  The
data available for comparison are all at low-energy, below isobar threshold.
They are fitted very well.  The predictions at higher-energy show structure at
isobar thresholds, which is very substantial for some partial waves.
However, the predictions of observables at these higher energies will require
the inclusion of partial waves with higher angular momenta.

In Section~VI we calculate the position and residues of the poles of the
boundary condition induced by the $\left[ q\left(1s_{1/2}\right)\right]^5$
$s\left(1s_{1/2}\right)$ quark configuration.  We discuss the implications
for the existence of observable exotics.
\goodbreak
\bigskip
\noindent{\bf II.\quad THE MESON EXCHANGE POTENTIAL}
\medskip
\noindent
We transform the meson-exchange potential of Ref.~[1] to configuration space.
The two-baryon, strangeness $-1$ channels considered are those of $\Lambda N$,
$\Sigma N$, $\Lambda\Delta$, $\Sigma\Delta$ and $\Sigma^*N$ [$\Sigma^*$
denotes the $\Sigma(1385)$ isobar].  All $J=0,1,$ and 2 partial waves are
considered.  In contrast to
the nucleon-nucleon system, in the $\Lambda N-\Sigma N$
sector there is coupling between the spin-singlet and spin-triplet partial
waves (of the same $J$ and parity) because of the significant mass
differences between the $N$, $\Lambda$ and $\Sigma$.  The meson exchanges
included are those of the $\pi$, $\sigma$ (correlated $2\pi$ exchange), $\rho$,
$\omega$, $K$, $K^*$, $\eta$, $\eta'$ and $\phi$ mesons.  The last three mesons
are omitted in Ref.~[1], but are included in Ref.~[9].  We prefer to include
them, with $SU(6)$ derived couplings, because we believe that they are not
excluded by the dispersion results$^{10}$ given their small contributions.
Indeed, the $\eta$ contributions in the $NN$ interaction are small and
consistent with all the $NN$ data.$^{2-5}$
The masses of the channel and exchange particles are given in Table I, and
Table~II contains the coupling constants used (the same as in Refs.~[1] and
[9]).  The choice of meson coupling is further discussed in Section~V.
The exchange diagrams are shown in Fig.~1.
\midinsert
$$\hbox{\vbox{\offinterlineskip
\def\superstrut{\hbox{\vrule height 15pt depth 10pt width 0pt}}
\def\strut{\hbox{\vrule height 8.5pt depth 3.5pt width 0pt}}
\hrule
\halign{
\strut\vrule#\tabskip 0.1in&
\hfil$#$\hfil &
\vrule#&
\hfil#&
\vrule#\tabskip 0.0in\cr
& \multispan 3\hfil{\bf Table I:}\hfil & \cr
& \multispan 3\hfil {Hadron Masses}\hfil &\cr\noalign{\hrule}
\superstrut& \omit\hfil Hadron\hfil && $\matrix{\hbox{Mass}\cr\hbox{(MeV)}\cr}$
 & \cr\noalign{\hrule}
& p && 938.3 & \cr
& n && 939.6 & \cr
& \Lambda && 1116.0 & \cr
& \Sigma && 1197.3 & \cr
& \Delta && 1232.0 & \cr
& \Sigma^* && 1385.0 & \cr\noalign{\hrule}
& \sigma && 550.0 & \cr
& \eta && 548.8 & \cr
& \eta' && 957.5 & \cr
& \pi && 135.0 & \cr
& \omega && 782.6 & \cr
& \phi && 1020.0 & \cr
& \rho && 770.0 & \cr
& K && 495.8 & \cr
& K^* && 895.0 & \cr\noalign{\hrule}}}}$$\endinsert

\pageinsert
$$\hbox{\vbox{\offinterlineskip
\def\superstrut{\hbox{\vrule height 12pt depth 5pt width 0pt}}
\def\strut{\hbox{\vrule height 8.5pt depth 3.5pt width 0pt}}
\hrule
\halign{
\strut\vrule#\tabskip 0.2in&
\hfil$#$\hfil &
\vrule#&
\hfil$#$&
\vrule#&
\hfil$#$&
\vrule#\tabskip 0.0in\cr
& \multispan 5\hfil{\bf Table II:} Coupling Constants\hfil &
\cr\noalign{\hrule}
& \multispan 5\hfil{\bf (a)}\hfil & \cr\noalign{\hrule}
\superstrut&
\hbox{Vertex} && g_i/\sqrt{4\pi} &&  & \cr\noalign{\hrule}
& NN\sigma && 2.385 && & \cr
& \Lambda\Lambda\sigma && 2.306 && & \cr
& \Sigma\Sigma\sigma && 3.061 && & \cr
& NN\eta && 2.730 && & \cr
& \Lambda\Lambda\eta && -1.360 && & \cr
& \Sigma\Sigma\eta && 2.560 && & \cr
& NN\eta' && 3.888 && & \cr
& \Lambda\Lambda\eta' && 4.650 && & \cr
& \Sigma\Sigma\eta' && 3.840 && & \cr
& NN\pi && 3.795 && & \cr
& \Lambda\Sigma\pi && 2.629 && & \cr
& \Sigma\Sigma\pi && 3.036 && & \cr
& N\Lambda K && -3.944 && & \cr
& N\Sigma K && 0.759 && & \cr\noalign{\hrule}
& \multispan5\hfil{\bf (b)}\hfil & \cr\noalign{\hrule}
\superstrut&
\hbox{Vertex} && g^v_i/\sqrt{4\pi} && g^t_i/\sqrt{4\pi} & \cr\noalign{\hrule}
& NN\omega && 4.472 && 0.0 & \cr
& \Lambda\Lambda\omega && 2.981 && -2.796 & \cr
& \Sigma\Sigma\omega && 2.981 && 2.796 & \cr
& NN\phi && -1.120 && -0.51 & \cr
& \Lambda\Lambda\phi && -1.960 && -4.30 & \cr
& \Sigma\Sigma\phi && -1.960 && 1.75 & \cr
& NN\rho && 0.917 && 5.591 & \cr
& \Lambda\Sigma\rho && 0.0 && 4.509 & \cr
& \Sigma\Sigma\rho && 1.834 && 3.372 & \cr
& N\Lambda K^* && -1.588 && -5.175 & \cr
& N\Sigma K^* && -0.917 && 2.219 & \cr\noalign{\hrule}
&\multispan5\hfil{\bf (c)}\hfil & \cr\noalign{\hrule}
\superstrut&\hbox{Vertex} && f^*_2/\sqrt{4\pi} && & \cr\noalign{\hrule}
& N\Delta\pi && 0.473 && & \cr
& \Lambda\Sigma^*\pi && 0.344 && & \cr
& \Sigma\Sigma^*\pi && -0.193 && & \cr
& N\Delta\rho && 4.522 && & \cr
& \Lambda\Sigma^*\rho && 3.198 && & \cr
& \Sigma\Sigma^*\rho && -1.846 && & \cr
& \Sigma\Delta K && -0.473 && & \cr
& N\Sigma^*K && -0.193 && & \cr
& \Sigma\Delta K^* && -4.522 && & \cr
& N\Sigma^*K^* && -1.846 && & \cr\noalign{\hrule}}}}$$\vfill\endinsert

In the $\Lambda N - \Sigma N$ sector the non-local potentials of Ref.~[1]
(without cut-offs) have
been expanded to bilinear order in ${\svec k}={\svec q} - {\svec q}'$ and
${\svec P} = - {1\over 2}\left( {\svec q} + q'\right)$, where $\pm{\svec q}$
and $\pm{\svec q}'$ are the initial and final three-momenta of the reaction
particles in the center-of-momentum frame (Fig.~2).  This results in the form
$( k = |{\svec k}|$ and $P = |{\svec P}|)$,
$$\eqalignno{
V\left( {\svec k},{\svec P}\right) &= V_c (k,P) {\tt 1\hskip -.27em l} +
V_\sigma (k,P) {\svec\sigma}_1\cdot {\svec\sigma}_2 + V_T (k,P) S_{12}  \cr
&+ V_{SL} (k,P) {i\over 2} \left( {\svec\sigma}_1+{\svec\sigma}_2\right)\cdot
\left({\svec k} \times {\svec P}\right) + V_{SLA} (k,P) {i\over 2} \left(
{\svec\sigma}_1-{\svec\sigma}_2\right) \cdot \left({\svec k} \times {\svec
P}\right) \cr
&+ V_{SSL} (k,P) Q_{12} \ \ ,&(2.1)\cr}$$
where
$$\eqalignno{S_{12}
&= 3\left( {\svec\sigma}_1 \cdot {\svec k}\right) \left({\svec\sigma}_2 \cdot
{\svec k}\right) -
k^2{\svec\sigma}_1 \cdot {\svec\sigma}_2\ \ ,&(2.2)\cr\noalign{\hbox{and}}
Q_{12} &= - \left( {\svec\sigma}_1 \times {\svec\sigma}_2\right) \cdot \left(
{\svec k} \times {\svec P}\right)\ \ .&(2.3) \cr}$$

Note that the last two operators of Eq.~(2.1) are odd in the exchange of
particles 1 and 2; therefore, the coefficients $V_{ SLA}$ and $V_{ SSL}$
must change sign when masses $M_1$ and $M'_1$ are exchanged with $M_2$ and
$M'_2$.  When the masses are the same, as in $NN$ scattering, these terms
vanish.  In addition, the operator $Q_{12}$ is odd under time reversal so that
$V_{ SSL}$ changes sign under the interchange of $M_1$ and $M_2$ with
$M'_1$ and $M'_2$.  The mass differences are relatively small, but these terms
are included because they are the only contributions to the mixing of singlet
and triplet states (such as ${}^1P_1$ and ${}^3P_1$).  Consequently,  a new
mixing parameter, $\epsilon^{ST}_J$ is required, in analogy to the tensor
mixing parameter $\epsilon_J$.

We note that to this order in $k$ and $P$ one more term arises, depending on
the differences between initial and
final masses but not odd under particle exchange.  We ignore this term as it
only occurs in transitions that also include terms not dependent on mass
differences.

In performing the Fourier transform with respect to ${\svec r}_1 = {\svec
r} - {\svec r}'$ and ${\svec r}_2 = {1\over 2} \left( {\svec r} +
{\svec r}'\right)$, where ${\svec r}$ and ${\svec r}'$ are the relative
coordinates in the initial and final state, respectively,
the coefficients $V_i (k,P)$
are expanded to the appropriate
order in $k^2$. That is, $V_c$ and $V_\sigma$ are expanded to ${\cal
O}(k^2)$, while $V_T$, $V_{SL}$, $V_{SLA}$ and $V_{SSL}$ are only needed to
${\cal O}(1)$ in $k^2$, because the operators are already quadratic in $k$ and
$P$.  Terms of ${\cal O}(P^2)$ are omitted in $V_c$ and $V_\sigma$ because it
has been found that explicit non-locality is very small for $r>1$~fm,$^{11}$
which is the range of $r$ for the potentials in this boundary condition model.

The result for the $\Lambda N-\Sigma N$ sector in configuration space, for the
exchange of a meson of mass $m_\alpha$, is of the form ($\hbar=c=1$),
$$\eqalignno{
V({\svec r}) &= \Biggl[ C^0_C + C^1_C + C_\sigma {\svec\sigma}_1\cdot
{\svec\sigma}_2 + C_T \left( 1 + {3\over m_\alpha r} + {3\over
\left(m_\alpha r\right)^2}
\right) S_{12} (\hat r) \cr
&+ C_{SL} \left( {1\over m_\alpha r} + {1\over \left( m_\alpha r\right)^2}
\right) {\svec L}\cdot {\svec S} + C_{SLA} \left( {1\over m_\alpha r} + {1\over
\left( m_\alpha r\right)^2}\right) {{\svec L}\over 2} \cdot
\left({\svec\sigma}_1-{\svec\sigma}_2\right) \cr
&+ C_{SSL} \left( {1\over m_\alpha r} + {1\over \left( m_\alpha r\right)^2}
\right) i \left( {\svec\sigma}_1\times {\svec\sigma}_2 \cdot {\svec L}
\right)\biggr] {e^{-m_\alpha r}\over m_\alpha r}\ \ . &(2.4)\cr}$$
Apart from the isospin dependence, presented later, the coefficients are as
follows.  For a scalar meson exchange ($\sigma$):
$$\eqalignno{
C^0_c &= - {g_1 g_2\over 4\pi} m_\alpha\ \ ,&(2.5a)\cr
4C^1_c &= - C_{SL} = {g_1 g_2\over 4\pi}\ {m^3_\alpha\over 4} \left( {1\over
M'_1M_1} + {1\over M'_2M_2}\right) \ \ ,&(2.5b)\cr
C_{SLA} &= - {g_1g_2\over 4\pi} \  {m^3_\alpha\over 4} \left( {1\over
M'_2M_2} - {1\over M'_1M_1}\right)\ \ ,&(2.5c)\cr
C_\sigma &= C_T = C_{SSL} = 0\ \ .&(2.5d)\cr}$$
For a pseudoscalar meson exchange $(\pi,\eta,\eta',K)$:
$$\eqalignno{C_\sigma &=C_T = {g_1g_2\over 4\pi}\ {m^3_\alpha\over 48} \left(
{1\over M'_1} + {1\over M_1}\right) \left( {1\over M'_2}+{1\over M_2}\right)\
\ ,&(2.6a)\cr
C_{SSL} &= - {g_1g_2\over 4\pi}\  {m^3_\alpha\over 16} \left[ \left( {1\over
M'_1} - {1\over M_1}\right) \left( {1\over M'_2} + {1\over M_2} \right) -
\left( {1\over M'_2} - {1\over M_2}\right) \left( {1\over M'_1} + {1\over
M_1}\right)\right] ,\qquad \quad &(2.6b)\cr
C^0_C &= C^1_C = C_{SL} = C_{SLA} = 0\ \ .&(2.6c)\cr}$$
For a vector meson exchange $(\rho,\omega,\phi,K^*)$:
$$\eqalignno{C^0_c &= {m_\alpha\over 4\pi} \left[ G_1 G_2 - {G_1 g^t_2\over 2M}
\left( M'_2 + M_2\right) - {G_2 g^t_1\over 2M} \left( M'_1 + M_1\right) +
{g^t_1g^t_2\over 4M^2}
\left( M'_1+M_1\right) \left( M'_2 + M_2\right)\right]\ \ ,\cr
&&(2.7a)
\cr\noalign{\vskip 0.3cm}
C^1_c &= {m^3_\alpha\over 4\pi} \Biggl[ {G_1G_2\over 16} \left\{ \left(
{1\over M_1} - {1\over M'_1}\right) \left( {1\over M_2}-{1\over M'_2}\right) -
{1\over M'_1M_1}-{1\over M'_2M_2}\right\} \cr
&- {G_1 g^t_2\over 16M} \left\{ {\left( M'_2+M_2\right)\over 2} \left( {1\over
M'_2 M_2} - {1\over M'_1 M_1}\right) + \left( {1\over M'_2}+{1\over M_2}
\right) \right\} \cr
&- {G_2 g^t_1\over 16M} \left\{ {\left( M'_1 + M_1\right)\over 2} \left(
{1\over M'_1M_1} - {1\over M'_2 M_2}\right) + \left( {1\over M'_1} + {1\over
M_1}\right) \right\} \cr
&+ {g^t_1g^t_2\over 32M^2} \left\{ {\left( M'_1 + M_1\right)\over 2}
\left(M'_2 + M_2\right) \left(
{1\over M'_1M_1} + {1\over M'_2 M_2}\right) \right. \cr
& + \left. \left( M'_1 + M_1\right) \left( {1\over M'_2} + {1\over
M_2}\right) + \left( M'_2 + M_2\right) \left( {1\over M'_1} + {1\over
M_1}\right) \right\} \Biggr]\ \ ,&(2.7b)\cr\noalign{\vskip 0.3cm}
C_\sigma &= - 2C_T = {m^3_\alpha\over 4\pi} \  {G_1G_2\over 24} \left( {1\over
M'_1} + {1\over M_1}\right) \left( {1\over M'_2} + {1\over M_2}\right) \ \
,&(2.7c)\cr\noalign{\vskip 0.3cm}
C_{SL} &= - {M^3_\alpha\over 32\pi} \Biggl[ 2G_1 G_2 \left( {1\over M'_1M_1}
+{1\over M'_2 M_2} + {2\over M'_2 M_1} + {2\over M'_1M_2}\right) \cr
&- {G_1 g^t_2\over M} \left( {M'_2 + M_2\over M'_1 M_1} + {2\over M_1} +
{2\over M'_1} - {1\over M_2} - {1\over M'_2}\right) \cr
&- {G_2 g^t_1\over M} \left( {M'_1+ M_1\over M'_2 M_2} + {2\over M_2} +
{2\over M'_2} - {1\over M_1} - {1\over M'_1}\right) \cr
&- {g^t_1 g^t_2\over 2M^2} \left( M'_1 + M_1\right) \left( M'_2 + M_2\right)
\left( {1\over M'_1M_1} + {1\over M'_2 M_2}\right) \Biggr]\ \
,&(2.7d)\cr\noalign{\vskip 0.3cm}
C_{SLA} &= - {m^3_\alpha\over 32\pi} \Biggl[ 2G_1 G_2 \left({1\over M'_1M_1} -
{1\over M'_2 M_2}\right)\cr
&- {G_1g^t_2\over M} \left( {M'_2 + M_2\over M'_1 M_1} + {2\over M_1} +
{2\over M'_1} + {1\over M_2} + {1\over M'_2} \right) \cr
&+ {G_2 g^t_1\over M} \left( {M'_1 +M_1\over M'_2M_2} + {2\over M_2} + {2\over
M'_2} + {1\over M_1} + {1\over M'_1} \right) \cr
&- {g^t_1 g^t_2\over 2M^2} \left( M'_1 + M_1\right) \left( M'_2 + M_2\right)
\left( {1\over M'_1M_1} - {1\over M'_2 M_2}\right) \Biggl]\ \
,&(2.7e)\cr\noalign{\vskip 0.3cm}
C_{SSL} &= {G_1 G_2\over 4\pi}   {m^3_\alpha\over 16} \left[ \left( {1\over
M'_1} - {1\over M_1}\right) \left( {1\over M'_2} + {1\over M_2} \right) -
\left( {1\over M'_2} - {1\over M_2} \right) \left( {1\over M'_1}+{1\over M_1}
\right)\right]\ \ . &(2.7f)\cr}$$

In the above coefficients the coupling constants are defined by the vertex
functions given in Appendix A of Ref.~[1].
The $g_i$ are the standard scalar and pseudoscalar
coupling constants of the meson-baryon-baryon vertex on leg $i$ (for example
$g^2_{NN\pi}/4\pi = 14.4$), the $g^t_i$ are the vector meson exchange tensor
coupling constants, and
$$G_i = g^v_i + {M'_1 + M_1\over 2M} g^t_i \ \ , \eqno(2.8)$$
where the $g^v_i$ are the vector meson exchange vector coupling constants.  $M$
is the nucleon mass.
The coupling constants $g_i$, $g^v_i$ and $g^t_i$ are given in Table~II, parts
(
a)
and (b).  Their determination is discussed in Section~V.

The above coefficients for the potential matrix components do not include
factors due to particle exchange and from isospin operators in the Lagrangian.
 For the strangeness exchange ($K,K^*$) diagrams the coefficients are
multiplied by $(-1)^{L+S}$, where $L$ is the angular momentum and $S$ the
total spin of each partial wave.  The vertex functions also have isospin and
transition isospin operators appropriate to the isospins of the meson and
baryons at the vertex.  The matrix elements of these operators lead to the
isospin factors of Table~III for each transition.  The operators and the
method of deriving the matrix elements are outlined in Appendix A.
The matrix elements of the non-transition
spin-spin, tensor and spin-orbit operators
are well-known.

As the transition potentials between the $\Lambda N - \Sigma N$ sector and the
channels containing $\Delta$ and $\Sigma^*$ isobars only contribute to
$\Lambda N - \Sigma N$ scattering in higher order, we keep only the lowest
order terms in $k$ and $P$ in the conversion to configuration space
potentials.  The diagrams producing an isobar always have one vertex requiring
the transition from spin-1/2 to spin-3/2 (both the $\Delta$ and $\Sigma^*$ are
spin-3/2).  The lowest order terms arise from pseudoscalar and vector meson
exchange.  Neglecting mass difference terms, both these exchanges lead to
spin-spin and tensor terms only.  We further simplify the isobar contributions
by keeping only the lowest mass mesons of the pseudoscalar and vector type,
non-strange and strange; {\it i.e.\/} the $\pi$, $\rho$, $K$ and $K^*$ as
shown in Fig.~1.  The result for any of the pseudoscalar meson
exchanges is of the form,
$$V_T^{PS} = f'_1 f^*_2 {m_\alpha\over 3} \left[ {\svec\sigma}_1 \cdot
{\svec S}_2 +
S^T_{12} \left( 1 + {3\over m_\alpha r} + {3\over \left( m_\alpha
r\right)^2}\right)\right] {e^{-m_\alpha r}\over m_\alpha r}\ \ , \eqno(2.9)$$
where ${\svec S}_2$ is the 1/2 to 3/2 transition spin operator and $S^T_{12} =
3\left( {\svec\sigma}_1\cdot \hat r\right) \left( {\svec S}_2\cdot \hat
r\right) - {\svec\sigma}_1 \cdot {\svec S}_2$.  For the pseudoscalar mesons
the $f'_1={g_1m_\alpha\over 2M_{\rm av}}$.  For the vector mesons,
$$V_T^V = f'_1 f^*_2 {m_\alpha\over 3} \left[ 2{\svec\sigma}_1\cdot {\svec S}_2
- S^T_{12} \left( 1 + {3\over m_\alpha r} + {3\over (m_\alpha r)^2}
\right)\right] {e^{-m_\alpha r}\over m_\alpha r}\ \ , \eqno(2.10)$$
where
$$f'_1 = {m_\alpha\over 2M_{\rm av}} \left( g^v_1 + g^t_1\right)\ \ .
\eqno(2.11)$$
The $f^*_2$ are the meson-baryon-isobar coupling constants;
they are tabulated
in Table II, part (c).  $M_{\rm av}$ is
the average of the two baryon masses at the vertex.  The choice of values is
 discussed in Section~V.
As before, the exchange phase (for $K$ and $K^*$ exchange) and the isospin
matrix elements are omitted in the above explicit description of the
potential, but are included in the calculations (see Table III and Appendix A).
 The transition spin-spin matrix elements are analogous to the transition
isospin-isospin matrix elements.  They are tabulated in Table~IV together with
the transition tensor matrix elements, whose derivation is discussed in
Appendix A.
\midinsert
$$\hbox{\vbox{\offinterlineskip
\def\superstrut{\hbox{\vrule height 15pt depth 10pt width 0pt}}
\def\strut{\hbox{\vrule height 8.5pt depth 3.5pt width 0pt}}
\hrule
\halign{
\strut\vrule#\tabskip 0.1in&
\hfil$#$\hfil&
\vrule#&
\hfil$#$\hfil&
\vrule#&
\hfil$#$\hfil&
\vrule#&
\hfil$#$\hfil&
\vrule#\tabskip 0.0in\cr
&\multispan7\hfil{\bf Table III:} Isospin Factors \hfil & \cr\noalign{\hrule}
&  &&  && \multispan3{\hfil Isospin Factors\hfil} &
\cr
& \hbox{Process} && \hbox{Mesons} && I = 1/2 &\omit&
I = 3/2 & \cr\noalign{\hrule}
& N\Lambda-N\Lambda && \sigma, \eta, \eta',\omega,\phi && 1 && 0 & \cr
& && \pi,\rho && 0 & & 0 & \cr
& && K,K^* && 1 && 0 & \cr\noalign{\hrule}
& N\Sigma-N\Sigma && \sigma,\eta,\eta'\omega,\phi && \phantom{-}1 && 1 & \cr
& && \pi,\rho && -2 && 1 & \cr
& && K,K^* && -1 && 2 & \cr\noalign{\hrule}
& N\Sigma-N\Lambda && \sigma,\eta, \eta',\omega,\phi && 0 && 0 & \cr
& && \pi, \rho&& \sqrt{3} && 0 & \cr
& && K,K^* && \sqrt{3} && 0 & \cr\noalign{\hrule}
& N\Lambda - \Delta\Sigma && \pi,\rho && - \sqrt{2} && 0 & \cr
& && K,K^* && 0 && 0 & \cr\noalign{\hrule}
& N\Sigma - \Delta\Sigma && \pi,\rho && - \sqrt{2/3} && - \sqrt{5/3} & \cr
& && K,K^* && - \sqrt{8/3} && \phantom{-}\sqrt{5/3} & \cr\noalign{\hrule}
& N\Sigma - \Delta\Lambda && \pi,\rho && 0 && 1 & \cr
& && K,K^* && 0 && 1 & \cr\noalign{\hrule}
& N\Lambda - N\Sigma^* && \pi,\rho && \sqrt{3} && 0 & \cr
& && K,K^* && \sqrt{3} && 0 & \cr\noalign{\hrule}
& N\Sigma - N\Sigma^* && \pi,\rho && -2 && 1 & \cr
& && K,K^* && -1 && 2 &\cr\noalign{\hrule}}}}$$\endinsert

\pageinsert
{\noindent\narrower{\bf Table~IV:}\quad
Transition spin-spin and transition tensor matrix
elements.  The
final states (bras) include one isobar, $S=3/2$.  The other final
state particle and  both initial state
particles (kets) have $S=1/2$.\smallskip}
$$\hbox{\vbox{\offinterlineskip
\def\strut{\hbox{\vrule height 13pt depth 7pt width 0pt}}
\hrule
\halign{
\strut\vrule#\tabskip 0.2in&
$#$\hfil&
\vrule#&
$#$\hfil&
\vrule#\tabskip 0.0in\cr
& \langle {}^5D_0|{\svec\sigma}_1 \cdot {\svec S}_2 |{}^1S_0\rangle = 0 &&
\langle {}^5D_0| S_{12}{}^T| {}^1S_0\rangle = - \sqrt{6} &
\cr\noalign{\hrule}
& \langle {}^3S_1| {\svec\sigma}_1\cdot{\svec S}_2| {}^3S_1\rangle = -
\sqrt{8/3} && \langle {}^3S_1| S_{12}{}^T|{}^3S_1\rangle = 0 &
\cr\noalign{\hrule}
& \langle {}^3S_1| {\svec\sigma}\cdot {\svec S}_2|{}^3D_1\rangle = 0 && \langle
{}^3S_1 | S_{12}{}^T | {}^3D_1\rangle=\sqrt{1/3} &
\cr\noalign{\hrule}
& \langle {}^3D_1| {\svec\sigma}_1\cdot{\svec S}_2|{}^3D_1\rangle =-
\sqrt{8/3} && \langle {}^3D_1| S_{12}{}^T|{}^3D_1\rangle = - \sqrt{1/6} &
\cr\noalign{\hrule}
& \langle {}^5D_1|{\svec\sigma}_1\cdot{\svec S}_2| {}^3 S_1\rangle = 0 &&
\langle {}^5D_1| S_{12}{}^T| {}^3S_1\rangle = \sqrt{3} &
\cr\noalign{\hrule}
& \langle {}^5D_1| {\svec\sigma}_1\cdot {\svec S}_2| {}^3D_1\rangle = 0 &&
\langle {}^5D_1|S_{12}{}^T| {}^3D_1\rangle = \sqrt{3/2} &
\cr\noalign{\hrule}
& \langle {}^3P_0|{\svec\sigma}_1\cdot {\svec S}_2|{}^3P_0\rangle
=-\sqrt{8/3} &&
\langle {}^3P_0|S_{12}{}^T| {}^3P_0\rangle =-\sqrt{2/3} & \cr\noalign{\hrule}
& \langle {}^3P_1|{\svec\sigma}_1\cdot{\svec S}_2|{}^1P_1\rangle = 0 && \langle
{}^3P_1| S_{12}{}^T | {}^1P_1\rangle = 0 & \cr\noalign{\hrule}
& \langle {}^3P_1 |{\svec\sigma}_1\cdot {\svec S}_2|{}^3P_1\rangle =
-\sqrt{8/3} && \langle {}^3P_1 | S_{12}{}^T| {}^3P_1\rangle = \sqrt{1/6} &
\cr\noalign{\hrule}
& \langle {}^5P_1 | {\svec\sigma}_1\cdot {\svec S}_1| {}^1P_1\rangle = 0 &&
\langle {}^5 P_1 | S_{12}{}^T | {}^1P_1\rangle = 2\sqrt{3/5} &
\cr\noalign{\hrule}
& \langle {}^5P_1| {\svec\sigma}_1\cdot {\svec S}_2|{}^3P_1\rangle = 0 &&
\langle {}^5 P_1| S_{12}{}^T | {}^3P_1 \rangle = 3\sqrt{3/10} &
\cr\noalign{\hrule}
& \langle {}^5F_1| {\svec\sigma}_1\cdot{\svec S}_2| {}^1P_1\rangle = 0 &&
\langle {}^5 F_1| S_{12}{}^T | {}^1P_1\rangle = - 3\sqrt{2/5} &
\cr\noalign{\hrule}
& \langle {}^5F_1|{\svec\sigma}_1\cdot{\svec S}_2|{}^3P_1\rangle = 0 &&
\langle{}^5F_1|S_{12}{}^T|{}^3P_1\rangle = 3\sqrt{1/5} &
\cr\noalign{\hrule}
& \langle {}^3P_2| {\svec\sigma}_1\cdot{\svec S}_2| {}^3P_2\rangle = -
\sqrt{8/3} && \langle {}^3P_2| S_{12}{}^T| {}^3P_2\rangle = - 1/(5\sqrt{6}) &
\cr\noalign{\hrule}
& \langle{}^3P_2|{\svec\sigma}_1\cdot{\svec S}_2| {}^3F_2\rangle = 0 && \langle
{}^3P_2 | S_{12}{}^T| {}^3F_2\rangle = 3/5 & \cr\noalign{\hrule}
& \langle{}^5P_2| {\svec\sigma}_1\cdot{\svec S}_2| {}^3P_2\rangle = 0 &&
\langle{}^5P_2| S_{12}{}^T|{}^3P_2\rangle = -(3/5)\sqrt{3/2} &
\cr\noalign{\hrule}
& \langle {}^5P_2| {\svec\sigma}_1\cdot{\svec S}_2| {}^3F_2\rangle = 0 &&
\langle {}^5P_2| S_{12}{}^T|{}^3F_2\rangle = - 3/5 & \cr\noalign{\hrule}
& \langle {}^3F_2| {\svec\sigma}_1\cdot{\svec S}_2| {}^3F_2\rangle = -
\sqrt{8/3} && \langle{}^3F_2| S_{12}{}^T | {}^3F_2\rangle=-(2/5)\sqrt{2/3} &
\cr\noalign{\hrule}
& \langle {}^5F_2| {\svec\sigma}_1\cdot{\svec S}_2|{}^3P_2\rangle = 0 &&
\langle {}^5 F_2| S_{12}{}^T | {}^3P_2 \rangle = (3/5) \sqrt{6} &
\cr\noalign{\hrule}
& \langle {}^5F_2| {\svec\sigma}_1\cdot {\svec S}_2| {}^3F_2\rangle = 0 &&
\langle {}^5 F_2 | S_{12}{}^T | {}^3F_2\rangle = 6/5 &
\cr\noalign{\hrule}}}}$$
\vfill
\endinsert
\goodbreak
\bigskip
\noindent{\bf III.\quad FORM OF THE BOUNDARY CONDITION}
\medskip
\nobreak
$R$-matrix theory$^{12}$
relates a boundary condition on the wave function external
to a closed boundary surface to the energies and wave functions of interior
eigenstates obeying a homogeneous boundary condition at the same surface.
This is a powerful tool if the physics is well-approximated by different
simple Hamiltonians and degrees of freedom in the
internal and external regions.  The
exact description of the systems is, of course, the same in all of space, but
the feasible approximation may differ importantly.  In nuclear reactions, to
which $R$-matrix methods were first applied, the exterior is well-represented
by a  Coulomb potential and the tail of the optical model potential.  In the
interior, where many-body excitations are easily produced, the level structure
of the compound nucleus is introduced.

In contemporary QCD we have a similar situation for particle
reactions.   Because of the confinement property of QCD, in the exterior
region (low momentum transfer) hadrons interact via the exchange of other
hadrons, describable by a potential.  In the interior, the QCD property of
asymptotic freedom (high momentum transfer) implies that the individual
 quark degrees of freedom are dominant and that the quark interactions
are perturbative.  Lattice QCD calculations$^8$ indicate that this
transition takes place over a small distance, so that the separating boundary
surface is given by a sphere of radius $R_0$ which is moderately
well-determined
by the QCD transition.

The interior dynamics is simpler than that of the compound nucleus.
In the simplest approximation the interior quark wave functions may be taken
as an antisymmetrized product of free Dirac wavefunctions satisfying the
condition on the boundary, $R_0$.  The energy of the quark configurations is
given by the sum of the Dirac energies, the perturbative one-gluon exchange
contribution, and the non-perturbative vacuum condensate energy (the
difference between the physical and perturbative vacua).  This is the bag
model type of Hamiltonian.  The parameters (quark masses, strong coupling
constant and condensate energy) can be taken from the fit of a bag model to
the hadron spectrum.

In principle, any homogeneous boundary condition on the interior wave
functions may be chosen.  The
confinement property requires that the quark wave functions vanish near $R_0$,
so that the most rapid convergence (in the sum over the complete set of
states) will be obtained with the condition $\psi_{\rm int}(R_0)=0$.  The bag
model boundary condition is not the vanishing of the wave function, but the
vanishing of the current.  However the wavefunction is zero at an
infinitesimally small distance beyond the radius at which the current
vanishes.  Therefore the bag model boundary conditions are appropriate.

If the two interacting hadrons are comprised of $n=n_1+n_2$ quarks, in the
exterior region all degrees of freedom are suppressed except for the relative
separation ${\svec r}$ of the two hadrons.  Averaging the free wave functions
over the remaining degrees of freedom, one obtains$^{13}$
the boundary radius in the relative hadron coordinate,
$$r_0 = 1.37 \left[ \left( n_1+n_2\right) /n_1 n_2\right]^{1/2} R_0
\ \ .\eqno(3.1)$$
The exterior boundary condition is$^4$
$$r_0 {d\psi^W_\alpha\over dr_0} = \sum_\beta f^{(W)}_{\alpha\beta}
\psi^W_\beta
(r_0)\ \ ,\eqno(3.2)$$
where $W$ is the barycentric energy, $\alpha,\beta$ denote the channels, and
the $f$-matrix
$$f_{\alpha\beta} (W) = f^0_{\alpha\beta} + \sum\limits_\beta
{\rho^i_{\alpha\beta} \over W-W_i} \ \ .\eqno(3.3)$$
The $W_i$ are the energies of the internal quark configurations and, using the
virial theorem$^{13}$, the residue matrix is
$$\rho^i_{\alpha\beta} = - r_0 {\partial W\over\partial r_0} \xi^i_\alpha
\xi^i_\beta \ \ ,\eqno(3.4)$$
where $\xi^i_\alpha$ is the fractional percentage coefficient of external
state $\alpha$ with respect to the quark configuration $i$.

For the two-baryon, strangeness $-1$
system the lowest energy quark configuration
is the $\left[q\left(1s_{1/2}\right)\right]^5 s\left(1s_{1/2}\right)$
where $q$ is an
$u$ or $d$ quark and $s$ is a strange quark.  Previous experience with the
nucleon-nucleon system$^{4,\,7}$ has shown that the MIT bag model is not
consistent with two-hadron data.  On the other hand, the Cloudy bag model
(CBM) is consistent with the available data, and predicts exotic structures in
the region $W=2.6-3.0$~GeV for which there is some evidence.$^{6,\,7,\,14}$
The
nucleon-nucleon data also indicated that $R_0\approx 0.85\,R_{eq}$ where
$R_{eq}$ is the six-quark equilibrium bag radius.  This is consistent with the
$R$-matrix theory requirement that $R_0$ be within the region that is
asymptotically free, but where the external interaction is closely given by
the hadron exchange potential. Although the $\Lambda N - \Sigma N$ data does
not extend to sufficiently high energy to verify the consistency, we will also
use $R_0 \approx 0.85\,R_{eq}$ in this paper.

The derivation of $R_{eq}$, $W_i (r_0)$ and $\xi_i$ is presented in
Section~VI, predicting the positions and widths of the exotic structures.  For
the fit to the available experimental data, with energies well-below the first
$f$-matrix pole, the energy-dependence due to the poles is negligible.  The
pole structure is, therefore, ignored in those fits and the fitted $f^0_{ij}$
actually represent
$$f^0_{\alpha\beta}({\rm eff}) = f^0_{\alpha\beta} + \sum_i
{\rho^i_{\alpha\beta} \over W_{\rm eff} - W_i}\ \ ,\eqno(3.5)$$
with $W_{\rm eff}$ being an energy in the experimental range.
\goodbreak
\bigskip
\noindent{\bf IV.\quad AMPLITUDES AND OBSERVABLES}
\medskip
\nobreak
The observables are expressed in terms of complex amplitudes which are formed
from the partial wave $S$-matrix.  The $S$-matrix is obtained by integrating
the Schr\"odinger equation from the boundary condition at $r_0$ to distances
at which the potentials are negligible.  There the wave functions can be
compared with the asymptotic form.  The many-channel Schr\"odinger equation is
equivalent to a coupled system of partial wave equations.  In this case the
coupling is between a set of partial waves of fixed total angular momentum,
parity and isospin.  There can be one or two (spin-coupled or tensor coupled)
$\Lambda$ channels, one or two $\Sigma$ channels and several isobar channels.
Using the channel labels $\alpha,\beta$, we have ($\hbar=c=1$):
$$- {1\over M^r_\alpha}\ {d^2 U_\alpha(r)\over dr^2} + {L_\alpha \left(
L_\alpha +1\right)\over M^r_\alpha r^2} U_\alpha (r) + \sum\limits_\beta
\left[ {M^r_\beta\over M^r_\alpha}\right]^{1/2} V_{\alpha\beta} (r) U_\beta
(r) = {k^2_\alpha\over M^r_\alpha} U_\alpha (r) \ \ ,\eqno(4.1)$$
where the modified partial wave $U_\alpha(r) = r\psi_\alpha(r)$,
$$M^r_\alpha = {M^1_\alpha M^2_\alpha\over M^1_\alpha +
M^2_\alpha}\qquad\hbox{and}\qquad
\left[ \left( M^1_\alpha\right)^2 + k^2_\alpha\right]^{1/2} + \left[ \left(
M^2_\alpha\right)^2 + k^2_\alpha\right]^{1/2} = W\ \ .\eqno(4.2)$$
The hermiticity of $V_{\alpha\beta} = V_{\beta\alpha}$ preserves unitarity.
The integration is started with the boundary conditions
$$r_0 {dU_\alpha\over dr_0} = \sum\limits_\beta f_{\alpha\beta} U_\beta (r_0)
\ \ ,\eqno(4.3)$$
where the symmetry $f_{\alpha\beta}=f_{\beta\alpha}$ is also required by
unitarity.

The above equations ignore the width of the isobar, their central masses being
understood in the above equations.  This is adequate for the present
application in which the fitted data is well-below the isobar channel
thresholds.  The approximation does however exaggerate the threshold effects
which we shall display for some partial waves.  The method of distributing the
isobar masses to account for the width is described in Ref.~[3].  This will be
used in future work when predicting observables in the isobar and exotic
energy range.

To calculate observables we used two sets of amplitudes and checked to
see that both sets gave the same results.  In particular, the first
set we used were the eight helicity amplitudes; the second set
were the invariant amplitudes.  Note that there are only five
independent amplitudes for the $NN$ system.  Explicitly,
the eight helicity amplitudes are:$^{1,\,15}$
$$\eqalignno{\phi_1&= {1\over {4ik}}\sum_{J=0}^{\infty}
\biggl\{(2J+1)\alpha_{J,0}+
[(J+1)\alpha_{J,+}  +  J\alpha_{J,-}  +\sqrt{J(J+1)}(\alpha_{-,+}^
 {J}  +  \alpha_{+,-}^{J})]\biggr\}P_J\ \ ,\cr
&&(4.4a)\cr
\phi_2 &= {1\over {4ik}}\sum_{J=0}^{\infty} \biggl\{-(2J+1)\alpha_{J,0} +
 [ (J+1)\alpha_{J,+}  +  J\alpha_{J,-}  +\sqrt{J(J+1)}(\alpha_{-,+}^
{J}  +  \alpha_{+,-}^{J})]\biggr\}P_J\ \ ,\cr
&&(4.4b)\cr
\phi_3 &= {1\over {4ik}}\sum_{J=0}^{\infty}\biggl\{(2J+1)\alpha_{J,1}  +
[J\alpha_{J,+}  +  (J+1)\alpha_{J,-}  -\sqrt{J(J+1)}(\alpha_{-,+}^
{J}   +  \alpha_{+,-}^{J})]\biggr\}\cr
&\qquad\qquad\quad\times \biggl\{P_J + {1-\cos\theta\over J(J+1)}P'_J\biggr\}
\ \ ,&(4.4c) \cr
\phi_4 &= {1\over {4ik}}\sum_{J=0}^{\infty}\biggl\{-(2J+1)\alpha_{J,1}  +
 [J\alpha_{J,+}  +  (J+1)\alpha_{J,-}  -\sqrt{J(J+1)}(\alpha_{-,+}^
{J}   +  \alpha_{+,-}^{J})]\biggr\}\cr
&\qquad\qquad\quad\times \biggl\{-P_J+ {1+\cos\theta\over
J(J+1)}P'_J\biggr\}\ \ ,&(4.4d) \cr
\phi_5 &= {1\over {4ik}}\sum_{J=0}^{\infty} \biggl\{(2J+1)\alpha_{1,0}^J  +
[ -(J+1)\alpha_{-,+}^J  +  J\alpha_{+,-}^J  +\sqrt{J(J+1)}(\alpha_{J,+}
  -  \alpha_{J,-})]\biggr\}\cr
&\qquad\qquad\quad\times \biggl\{{P'_J \sin\theta\over \sqrt{J(J+1)}}\biggr\}
\ \ , &(4.4e)\cr
\phi_6 &= {1\over {4ik}}\sum_{J=0}^{\infty} \biggl\{-(2J+1)\alpha_{0,1}^J  +
[ -J\alpha_{-,+}^J  +  (J+1)\alpha_{+,-}^J  -\sqrt{J(J+1)}(\alpha_{J,+}
  -  \alpha_{J,-})]\biggr\}\cr
&\qquad\qquad\quad \times \biggl\{{P'_J \sin\theta\over \sqrt{J (J+1)}}\biggr\}
\ \ , &(4.4f)\cr
\phi_7 &= {1\over {4ik}}\sum_{J=0}^{\infty}\biggl\{(2J+1)\alpha_{1,0}^J  -
[ -(J+1)\alpha_{-,+}^J  +  J\alpha_{+,-}^J  +\sqrt{J(J+1)}(\alpha_{J,+}
  -  \alpha_{J,-})]\biggr\}\cr
&\qquad\qquad\quad \times \biggl\{{P'_J \sin\theta\over \sqrt{J(J+1)}}\biggr\}
\ \ ,&(4.4g)\cr
\phi_8 &= {1\over {4ik}}\sum_{J=0}^{\infty}\biggl\{-(2J+1)\alpha_{0,1}^J  -
[ -J\alpha_{-,+}^J  +  (J+1)\alpha_{+,-}^J  -\sqrt{J(J+1)}(\alpha_{J,+}
  -  \alpha_{J,-})]\biggr\}\cr
&\qquad\qquad\quad
\times \biggl\{{P'_J \sin\theta\over \sqrt{J(J+1)}}\biggr\}\ \ ,&(4.4h)
\cr}$$
where $P_J$ are the Legendre polynomials of order $J$, $P'_J$ are
derivatives of $P_J$ with respect to their argument, $\cos\theta$,
$\theta$ is the center of mass scattering angle, and $k$ is the
center-of-mass momentum.
When the final masses are the same as the initial masses,
then $\phi_8=-\phi_7$ and $\phi_6=-\phi_5$, leaving only six independent
complex amplitudes.
The diagonal partial wave amplitudes are:
$$\eqalignno{\alpha_{J,0} &= \hbox{spin coupled singlet}\
({}^1S_0,\ {}^1P_1,\ {}^1D_2,\
\hbox{\it etc.})\ \ ,&(4.5a) \cr
\alpha_{J,1} &= \hbox{spin coupled triplet}\ ({}^3P_1,{}^3D_2,
\ \hbox{\it etc.}) \ \ , &(4.5b)\cr
\alpha_{J,\pm} &= L =J\pm 1 \hbox{ coupled triplet}\ ({}^3P_0,
{}^3S_1-{}^3D_1,\ {}^3P_2-{}^3F_2,\ \hbox{\it etc.})\ \ .&(4.5c) \cr}$$
We note that the ${}^1S_0$ and the ${}^3P_0$ lack a coupled partner.
The $\alpha^J$ are off-diagonal amplitudes for coupled states.  Namely
$\alpha_{-,+}^J$ and $\alpha_{+,-}^J$ are for $^3(J-1)_J$ $\rightarrow$
$^3(J+1)
_J$ and $^3(J+1)_J$ $\rightarrow$ $^3(J-1)_J$ reactions and
$\alpha_{0,1}^J$ and $\alpha_{1,0}^J$ are for $^1J_J$ $\rightarrow$ $^3J_J$
 and $^3J_J$ $\rightarrow$ $^1J_J$.
The partial wave amplitudes can be constructed from the phase shift
parameters.  Namely,
$$\vbox{\eqalignno{\alpha_{J,0} &= \cos 2\epsilon^{01}_J \exp( 2i
\delta_{J,J,0} )
 -  \delta_{AB}
\ \ ,&(4.6a)\cr
\alpha_{J,1} &= \cos 2\epsilon^{10}_J\exp( 2i\delta_{J,J,1} )
-  \delta_{AB}\ \ ,&(4.6b)\cr
\alpha_{J,-} &=  \cos2\epsilon^{-+}_J \exp( 2i\delta_{J,J-1,1} ) -
\delta_{AB}\ \ ,
&(4.6c)\cr
\alpha_{J,+}&= \cos 2\epsilon^{+-}_J \exp \left(
2i\delta_{J,J+1,1}\right)-\delta_{AB}
\ \ ,&(4.6d)\cr
\alpha^J_{-,+} &= i \sin 2\epsilon^{-+}_J \exp \left[ i \left(
\delta_{J,J+1,1} +
\delta_{J,J-1,1} + \phi^{-+}_J\right)\right]\ \ ,&(4.6e)\cr
\alpha^J_{+,-} &= i \sin 2\epsilon^{+-}_J \exp \left[ i \left(
\delta_{J,J+1,1} +
\delta_{J,J-1,1} + \phi^{+-}_J\right)\right]\ \ ,&(4.6f)\cr
\alpha^J_{0,1} &= i \sin 2\epsilon^{01}_J \exp \left[ i \left( \delta_{J,J,0} +
\delta_{J,J,1} + \phi^{01}_J \right)\right]\ \ ,&(4.6g)\cr
\alpha^J_{1,0} &= i \sin 2\epsilon^{10}_J \exp \left[ i \left( \delta_{J,J,0} +
\delta_{J,J,1} + \phi^{10}_J\right)\right]\ \ ,&(4.6h)\cr}}$$
where $A,B$ represent $\Lambda N$ or $\Sigma N$ channels and
the complex phase shifts are denoted by $\delta_{J,L,S}$.
For $\Lambda N$ scattering, the $\epsilon$'s and $\delta$'s are real
and $\phi$ = 0 until $\Sigma$ threshold is reached.  Above threshold
and for $\Sigma N$ scattering at any energy the $\delta$'s are complex,
with $\Im\delta > 0.$  In our figures
 we use the notation that
$\delta\left({}^{2s+1}L_J\right) \equiv\Re \delta_{J,L,S}$,
$\eta\left( {}^{2s+1}L_J\right) \equiv \cos
2\epsilon^{\alpha\beta}_J\exp (-2\Im\delta_{J,L,S})$, and $\sin
2\bar{\epsilon}^{\alpha\beta}_J = \sin 2\epsilon^{\alpha\beta}_J \exp \left( -
\Im \delta_{J,\alpha} - \Im \delta_{J,\beta}\right)$ where $\alpha,\beta$ stand
for $0,1,+$ or $-$ as appropriate.
The additional parameters, $\phi$, are required to complete the
16 parameters needed to express the
eight complex amplitudes.  When the final masses are the same as the initial
masses then $\epsilon^{01}_J = \epsilon^{10}_J$, $\phi^{01}_J = \phi^{10}_J$,
$\epsilon^{-+}_J = \epsilon^{+-}_J$ and $\phi^{-+}_J = \phi^{+-}_J$, leaving
12 independent parameters to represent the six independent complex amplitudes.
The coupled channel parameters obey a sub- unitarity inequality because of
their coupling to other particle channels.

The invariant amplitudes are given in terms of the helicity amplitudes as:
$$\vbox{\eqalignno{a &= {1\over 2}\left[( \phi_1 + \phi_2 + \phi_3 - \phi_4 )
\cos\theta - (\phi_5 - \phi_6 - \phi_7 + \phi_8 )\sin\theta\right]\ \ ,
&(4.7a)\cr
b &= {1\over 2}(\phi_1 - \phi_2 + \phi_3 + \phi_4)\ \ ,&(4.7b)\cr
c &= {1\over 2}(-\phi_1 + \phi_2 + \phi_3 + \phi_4)\ \ ,&(4.7c)\cr
d &= {1\over 2}(\phi_1 + \phi_2 - \phi_3 + \phi_4)\ \ ,&(4.7d)\cr
e &= {-i\over 2}\left[(\phi_1 + \phi_2 + \phi_3 - \phi_4)
\sin\theta + (\phi_5 - \phi_6 - \phi_7 + \phi_8 )\cos\theta\right]
\ \ ,&(4.7e)\cr
f &= {i\over 2}(\phi_5 - \phi_6 + \phi_7 - \phi_8 )\ \ ,&(4.7f)\cr
g &= {1\over 2}(\phi_5 + \phi_6 - \phi_7 - \phi_8 )\ \ ,&(4.7g)\cr
h &= {1\over 2}(\phi_5 + \phi_6 + \phi_7 + \phi_8 )\ \ .&(4.7h)\cr}}$$
When the final masses are the same as the initial masses, $g=h=0$.

Note that in the $NN$ case, $\alpha_{0,1}^J$ and $\alpha_{1,0}^J$ do not
exist and $\alpha_{+,-}^J = \alpha_{-,+}^J$.  Then, it follows that
$\phi_5 = -\phi_7$ in addition to $\phi_7 = -\phi_8$, and $\phi_5 = -\phi_6$.
  Therefore,
$f = g = h = 0$ and there are only five independent amplitudes, as noted
above.

The differential cross section is given by
$${d\sigma \over d\Omega} = {1\over 2}\ {k' \over k} \sum_{i=1}^8
\left| \phi_i \right| ^2\ \ \ .\eqno(4.8)$$
In terms of the invariant amplitudes,
$${d\sigma \over d\Omega} = {1\over 2}\ {k' \over k}
\biggl\{\mid a \mid ^2 + \mid b \mid ^2 + \mid c \mid ^2 + \mid d \mid ^2
 + \mid e \mid ^2 + \mid f \mid ^2 + \mid g \mid ^2 + \mid h \mid ^2
\biggr\}\ \ , \eqno(4.9)$$
where $k$ is the magnitude of the center-of-mass  momentum of each particle in
the initial state and $k'$ is that of the final state.

For $\Sigma^+ p \rightarrow \Sigma^+ p$ and $\Lambda p \rightarrow \Lambda
p$, which are states of definite isospin, the amplitudes can be calculated
as above.  For states of mixed isospin, such as $\Sigma^- p, \Sigma^0 n$,
{\it etc.\/}, the amplitudes which go into calculating the differential cross
sections  are linear
combinations of the amplitudes of pure isospin.  In particular, if $M$ is
an amplitude such that $\langle I = {3 \over 2} \mid M \mid I = {3 \over 2}
\rangle = M_{3 \over 2}$  and $\langle I = {1 \over 2} \mid M \mid
 I = {1 \over 2}\rangle = M_{1 \over 2}$ then,
$$\eqalignno{\langle \Sigma^- p \mid M \mid \Sigma^- p \rangle &=
{1\over 3}M_{3\over 2} + {2\over 3}M_{1\over 2}\ \ ,&(4.10a)\cr
\langle \Sigma^- p \mid M \mid \Sigma^0 n \rangle &= {\sqrt{2}\over 3}
 \left(M_{3\over 2} - M_{1\over 2}\right)\ \ ,&(4.10b)\cr
\langle \Sigma^- p \mid M \mid \Lambda n \rangle &= {\sqrt{2\over 3}}M_{1\over
2}\ \ .&(4.10c)\cr}$$

Another important quantity is the total cross section into all available
channels.  This quantity is given by the optical theorem:
$$\sigma = {2\pi \over k}\Im\biggl\{\phi_1(0^{\circ}) +
\phi_3(0^{\circ})\biggr\} = {2\pi \over k}\Im\biggl\{a(0^{\circ}) +
b(0^{\circ})\biggr\}\ \ . \eqno(4.11)$$
\goodbreak
\bigskip
\noindent{\bf V.\quad METHOD AND RESULTS}
\medskip
\nobreak In this paper, we use a standard partial wave decomposition of the
interaction.  As we are concerned primarily with fitting the low energy
scattering data of the hyperon-nucleon interaction, we include only those
partial waves with $J\leq 2$.  Including partial waves with $J > 2$
introduces essentially no new parameters since, because of the angular
momentum barrier at these energies,
these waves will not be very sensitive to the boundary
condition at the core radius.
We have constructed three models of varying degrees of complexity.  In the
first model, only one meson exchange between the initial and final states
is considered.  The second model includes, in addition, the possibility of
intermediate $\Delta$ and $\Sigma^*$ isobar states.  The third model,
finally, includes the effects of the Coulomb force.  To be specific, the
baryons and mesons used in our models, with their masses, are shown in
Table I.

As usual, the scalar $\sigma$ meson should be thought of as an effective
parameterization of correlated 2$\pi$ exchange.  The parameters of this
model are the vertex coupling constants and the constant $f$-matrix elements
which determine the boundary condition on the two body wavefunction.  The
coupling constants for the $NN$ and $N\Delta$ vertices with the $\pi$, $\rho$
and $\omega$ mesons are taken to be
those of the Bonn $NN$ potential.  All others with $\pi$, $\rho$, $\omega$,
$K$ and $K^*$ mesons and all vertices with $\eta,\eta'$ and $\phi$ mesons
are assumed to be fixed by
$SU(6)$ flavor $\times$ spin symmetry.  The values of these coupling
constants can be found in Refs.~[1] and [9]
and in Table~II.  Representative examples of the resulting potential
contributions are displayed in Figs.~3 and 4.
Figure 3 shows the contribution for $\Sigma N\to\Sigma N$ scattering of each
meson exchange to the six types of potentials in Eq.~(2.1).  It is
representative of their importance, when allowed, for $\Lambda N\to \Lambda N$
and $\Sigma N\to \Lambda N$ scattering.  The largest terms are the central
$\sigma$ and $\omega$ potentials, which are prohibited by isotopic spin in
$\Sigma N\leftrightarrow\Lambda N$ scattering.  Next in importance are the
pion tensor and $\rho$-meson central and spin-spin potentials.  The $\omega$
and $\rho$ mesons contribute substantial spin-orbit terms.  The two
antisymmetric spin-orbit potential terms are small as expected. The
strangeness-exchange potentials ($K$ and $K^*$) are small in Fig.~3 but are
more substantial in $\Sigma N\leftrightarrow \Lambda N$ scattering, and
important in $\Lambda N\to \Lambda N$ scattering where components are as large
as 10~MeV at 1.1~fm.  Figure 4 illustrates the meson exchange strengths for
isobar production in the $\Sigma N\to\Lambda\Delta$ case. The tensor
components of each meson exchange are most important.  Tensor terms are also
dominant for all of the other $\Lambda N-\Sigma N\to \Sigma\Delta
-\Lambda\Delta - \Sigma^* N$ couplings.

We now turn our attention to the determination of the $f$-matrix elements for
each partial wave.  We choose $r_0 = 1.082$~fm, which will be seen in Section
VI to correspond to between 0.80 and 0.85 of the equilibrium radius of the CBM
for all the $\left[q(1s_{1/2})\right]^5 s\left(1s_{1/2}\right)$ quark exotics.
The obvious place to start is with the $I = {3\over 2}$,
 $\Sigma^+ p \rightarrow \Sigma^+ p$ reaction since only one isospin enters
in calculation of the cross sections.  The diagonal elements of both the
$^1S_0$ and the $^3S_1$ waves were adjusted to fit the total cross section
data.  Then, we choose to increase the contribution from the $^1P_1$ wave
in order to obtain an angular distribution which fits the experimental data
.  Varying the $P$ wave parameters, while changing the shape of the
differential
cross section greatly, does not affect the total cross section very much at
all.  All other diagonal $f$-matrix elements are set to their neutral
values $L+1$, which corresponds to a free wave function for $r > r_0$ at
threshold.
 The off-diagonal $f$-matrix elements are set to zero.  This prescription
gives,
for the only non-zero or non-neutral $f$'s,

\baselineskip 12pt
$$\eqalignno{
f_{^1S_0}&=\bordermatrix{&\Sigma N (^1S_0)\cr \Sigma N (^1S_0)&1.162\
 (1.068)\cr}\ \ ,&(5.1a)\cr\noalign{\vskip 0.2cm}
f_{SD}&=\bordermatrix{&\Sigma N (^3S_1)&\Sigma N (^3D_1)\cr \Sigma N (
^3S_1)&1.850 (1.69)&0.0\cr \Sigma N (^3D_1)&&3.0\cr} \ \ ,&(5.1b)
\cr\noalign{\vskip 0.2cm}
f_{P}&=\bordermatrix{&\Sigma N (^1P_1)&\Sigma N (^3P_1)\cr \Sigma N (
^1P_1)&0.86 (0.75)&0.0\cr \Sigma N (^3P_1)&&2.0\cr} \ \ ,&(5.1c)\cr}$$

\baselineskip 24pt plus 2pt minus 2pt
\noindent
where the first values hold for Models 2 and 3, the values in
parenthesis are those for the no isobar/no Coulomb Model
1, when they differ.  Coupling to the isobar channels is accomplished through
 $\pi$, $K$, $\rho$, and $K^*$ exchange.  The open channels for
$I={3\over 2}$ are $\Sigma
\Delta$, $\Sigma^* N$, and $\Lambda \Delta$.  The highest energy threshold
$\Sigma^*\Delta$ is ignored.  There is both a spin-spin and
a tensor part to these interactions, as in Section II.

Both of these parameter sets yield scattering lengths of $a_s = -4.60~$fm
and $a_t = 0.33$~fm for Models 1 and 2.  Model 3, which is just Model 2
with Coulomb effects, uses the same parameters as Model 2, and gives
scattering lengths of $a_s = -3.78$~fm and $a_t = 0.35$~fm.

Table~V is a comparison of scattering lengths from this analysis,
Holzenkamp's {\it et al.\/}
 (Ref.~[1]), and the Nijmegen group (Ref.~[9]).  Using the notation in Ref.~[1]
$$\hbox{\vbox{\offinterlineskip
\def\strut{\hbox{\vrule height 10pt depth 5pt width 0pt}}
\hrule
\halign{
\strut\vrule#\tabskip 0.2in&
\hfil#\hfil &
\vrule#&
\hfil$#$\hfil &
\vrule#&
\hfil$#$\hfil &
\vrule#\tabskip 0.0in\cr
&\multispan5\hfil{\bf Table V:} $\Sigma^+p$ Scattering Lengths\hfil &
\cr\noalign{\hrule}
& && a_s && a_t & \cr\noalign{\hrule}
& Model 1 && -4.60 && \phantom{-}0.33 & \cr
& Model 2 && -4.60 && \phantom{-}0.33 & \cr
& Model 3 && -3.78 && \phantom{-}0.35 & \cr
& $A$ && -2.28 && -0.78 & \cr
& $B$ && -1.10 && -0.90 & \cr
& $D$ (Nijmegen) && -4.61 && \phantom{-}0.32 & \cr
& $F$ (Nijmegen) && -3.84 && \phantom{-}0.62 & \cr\noalign{\hrule}}}}$$

The low-energy results of our Models 1, 2, and 3 are shown in Fig.~5
against the experimental data of the Heidelberg group.$^{16}$ The solid curve,
dashed curve, and dotted curve correspond to Models 3, 2, and 1,
respectively.
As can be seen, the effects of turning on the Coulomb effects are of the
order of about 10 percent, being more significant at small angle and low
energies. The total cross section was obtained by numerically integrating the
calculated differential cross section.  These values agree with those demanded
by the optical theorem to within a few tenths of one percent.

During this fitting, the effect of varying the $f$-matrix elements for
different partial waves on the angular
distribution was examined.  The results of this investigation agree with
those found previously:$^{9}$ (a) a strong dependence on the $^1P_1$ wave; (b)
almost no dependence on $^3P_0$; (c) almost no dependence on $^3P_1$;  (d)
slight dependence on the coupled $3P_2$ wave.  The large peak in
the forward direction is evidence for a strong contribution from the $P$
waves.

Finding the $f$-matrix elements for the $I=1/2$ partial waves was more
tedious.  As before, the diagonal elements of the $s$-waves were fitted to
reproduce the $\Sigma^-p\to \Sigma^-p$ total cross section. As the $I=1/2$
channel also couples to the $\Lambda N$ channel, we must simultaneously fit
the $\Lambda p\to \Lambda p$ and $\Sigma^-p\to\Lambda n$ data.  We found it
necessary to adjust the parameters which most directly affect the $\Lambda$
production, $f_{{}^1S_0}$ [$\Sigma N\to \Lambda N$] and $f_{SD}$ [$\Sigma
N\left({}^3S_1\right)\to \Lambda N \left( {}^3 S_1\right)$].  In addition, to
fit the high energy $\Lambda p\to \Lambda p$ data we found that we had to
increase the attraction caused by coupling to the isobar channels.  This
$f$-matrix coupling was made to the $\Sigma^*N$ channels, as this has the
lowest threshold of the isobar channels.  Lastly, the contribution from all
of the $p$-waves was increased, by adjusting the $f$-matrix, in order to fit
the forward-peaked angular distributions.  Once again, all other $f$-matrix
elements are set either to their neutral values, $2L+1$ on the diagonal, or to
zero. Then for the non-trivial $f$'s we find,

\baselineskip 12pt
$$\eqalignno{
f_{^1S_0}&=\bordermatrix{&\Sigma N (^1S_0)&\Lambda N (^1S_0)\cr \Sigma N (
^1S_0)&0.600 (0.466)&1.1 (0.9)\cr \Lambda N (^1S_0)&&1.245 (1.000)\cr}
\ \ ,&(5.2a)\cr\noalign{\vskip 0.2cm}
f_{^3P_0}&=\bordermatrix{&\Sigma N (^3P_0)&\Lambda N (^3P_0)\cr \Sigma N (
^3P_0)&2.0&3.0\cr \Lambda N (^3P_0)&&2.4\cr}\ \ ,
&(5.2b)\cr\noalign{\vskip 0.2cm}
f_{SD}&=\bordermatrix{&\Sigma N (^3S_1)&\Sigma N (^3D_1)&\Lambda N (^3S_1)
& \Lambda N (^3D_1)\cr \Sigma N (^3S_1)&1.180 (1.120)&0.0&0.4&0.0\cr
\Sigma N (^3D_1)&&3.0&0.0&0.0\cr \Lambda N (^3S_1)&&&5.91 (0.830)&0.0\cr
\Lambda N (^3D_1) &&&&3.0\cr}\ \ ,&(5.2c)\cr}$$
Also, for models 2 and 3, $f\left[ \Lambda N\left({}^3S_1\right) -
\Sigma^*N\left( {}^3S_1\right)\right]=-4.0$ the only non-zero or non-neutral
$f$-matrix component in the isobar sector.
$$\eqalignno{f_{PF}
&=\bordermatrix{&\Sigma N (^3P_2)&\Sigma N (^3F_2)&\Lambda N (^3P_2)
& \Lambda N (^3F_2)\cr \Sigma N (^3P_2)&2.0&0.0&1.3&0.0\cr
\Sigma N (^3F_2)&&4.0&0.0&0.0\cr \Lambda N (^3P_2)&&&2.0&0.0\cr
\Lambda N (^3F_2) &&&&4.0\cr}\ \ ,&(5.2d)\cr\noalign{\vskip 0.2cm}
f_{P}&=\bordermatrix{&\Sigma N (^1P_1)&\Sigma N (^3P_1)&\Lambda N (^1P_1)
& \Lambda N (^3P_1)\cr \Sigma N (^1P_1)&2.0&0.0&0.0&0.0\cr
\Sigma N (^3P_1)&&2.0&0.0&1.0\cr \Lambda N (^1P_1)&&&2.0&0.0\cr
\Lambda N (^3P_1) &&&&2.0\cr}\ \ .&(5.2e)\cr}$$

\baselineskip 24pt plus 2pt minus 2pt
\noindent
Again, the first values correspond to Models 2 and 3, the second
values for Model 1.  The coupling to the isobar channels is still
accomplished through the exchange of $\pi$, $K$, $\rho$, and $K^*$ mesons.
The open channels now are only the $\Sigma \Delta$ and the $\Sigma^* N$.

These parameter sets then yield the scattering lengths shown in Table VI:
$$\hbox{\vbox{\offinterlineskip
\def\strut{\hbox{\vrule height 10pt depth 5pt width 0pt}}
\hrule
\halign{
\strut\vrule#\tabskip 0.1in&
\hfil#\hfil&
\vrule#&
\hfil#\hfil&
\vrule#&
\hfil#\hfil&
\vrule#\tabskip0.0in\cr
&\multispan5\hfil{\bf Table VI:} $I=(1/2)\ \Sigma N$ Scattering Lengths\hfil
& \cr\noalign{\hrule}
& && $a_s$ (fm) && $a_t$ (fm) & \cr\noalign{\hrule}
& Model 1 && $-1.05$ -- $1.70\,i$ && $-3.18$ -- $1.82\,i$ & \cr
& Model 2 && $-0.90$ -- $1.97\,i$ && $-3.20$ -- $1.98\,i$ & \cr
& Model 3 && $-0.60$ -- $2.09\,i$ && $-3.13$ -- $2.28\,i$ &
\cr\noalign{\hrule}}}}$$

The theoretical predictions of our model for the differential and total
cross sections of the reactions $\Sigma^- p \rightarrow \Sigma^- p$,
$\Sigma^- p \rightarrow \Sigma^0 n$, $\Sigma^- p \rightarrow \Lambda n$ are
compared with experimental data$^{16,\,17}$ in Fig.~6.

The total cross sections for these three reactions were obtained by
integrating the angular distributions.  The optical theorem
relates, in this case, the sum of these three total cross sections to the
imaginary part of the forward scattering amplitude.  This relation was
checked and found to agree to within a few tenths of one percent.  Isospin
symmetry between the $\Sigma^-$ and $\Sigma^0$ particles is
broken by multiplying the cross sections by the kinematical
factor ${k'/ k}$.

Another interesting quantity is the inelastic capture ratio at rest as
defined in deSwart and Dullemond:$^{18}$
$$R = {1\over 4}\ {\sigma_s(\Sigma^-p\to \Sigma^0n) \over \sigma_s (\Sigma^- p
\to \Sigma^0n) + \sigma_s(\Sigma^-p\to \Lambda n)} + {3\over 4} \
{\sigma_t(\Sigma^- p\to\Sigma^0n)\over \sigma_t (\Sigma^-p\to\Sigma^0n) +
\sigma_t (\Sigma^- p\to\Lambda n)}\ \ .\eqno(5.3)$$
The experimental values are
$$R = \cases{0.470 \pm 0.003 & Ref.~[17] \cr
0.474\pm 0.016 & Ref.~[19]\cr}\ \ .\eqno(5.4)$$
In the present work we predict
$$R = \cases{ 0.474 & for Model 1 \cr
0.467 & for Model 2 \cr
0.597 & for Model 3 \cr}\ \ . \eqno(5.5)$$
It should be noted that the magnitude of the Coulomb effects in the
$\Sigma^- p \rightarrow \Sigma^0 n$ reaction is an upper bound since we did
not split the $\Sigma^-$ and $\Sigma^0$ channels
in our code.  Therefore the Coulomb
attraction is acting in the final as well as the initial state.  This
false Coulomb enhancement increases the ratio $R$.
This results in a strong dependence of $R$ on energy as threshold is
approached.  To minimize the spurious effect while staying near enough to
threshold we use $T(\hbox{lab}) = 0.5$~GeV.
Judging from the low energy predictions of Models 2 and 3, we should expect
the ratio $R$ to be in good agreement with the data if the charge symmetry
of the $\Sigma^-$ and $\Sigma^0$ were broken correctly.  It is also
important to note that the correct $\Sigma$ and $N$ masses were used to
calculate $R$.  That is, the $\Sigma^- p$ threshold is at $2135.6$~MeV/c$^2$.
  In all other cases, the masses given at the beginning of this
section are used.

As above, the effect of the different $P$ waves on the phase
parameters was investigated and the following results were obtained:  the
phases have (a) a moderate dependence on the $^1P_1$ wave; (b) almost no
dependence on $^3P_0$; (c) slight dependence on $^3P_1$; (d) strong
dependence on the coupled $^3P_2$ wave.

The calculated $\Lambda p \rightarrow \Lambda p$ and the $\Lambda
p\to\Sigma^0p$ total cross sections are
compared with the Rehovoth--Heidelberg$^{20}$ and Maryland$^{21}$
data in Fig.~7.
Because the $\Lambda$ has no charge, there are no Coulomb interactions in
the $\Lambda p$ system.  Therefore, there is no distinction between Models 2
and 3.  We use the notation of a solid curve to indicate that Model 2 is
the full model.
The $f$-matrices for the $\Lambda p$ partial waves are given by simply
exchanging the $\Lambda$ and $\Sigma$ columns in the matrices given above.
In comparison with other theoretical models and experimental data, the
above parameter sets yield the scattering lengths shown in Table~VII.
$$\hbox{\vbox{\offinterlineskip
\def\strut{\hbox{\vrule height 10pt depth 5pt width 0pt}}
\hrule
\halign{
\strut\vrule#\tabskip 0.2in&
\hfil#\hfil &
\vrule#&
\hfil$#$\hfil &
\vrule#&
\hfil$#$\hfil &
\vrule#\tabskip 0.0in\cr
&\multispan5\hfil{\bf Table VII:} $\Lambda N$ Scattering Lengths\hfil
& \cr\noalign{\hrule}
& && a_s\ \hbox{(fm)} && a_t\ \hbox{(fm)} & \cr\noalign{\hrule}
& Model 1 && -2.97 && -1.92 & \cr
& Model 2 && -2.98 && -1.70 & \cr
& $A$ && -1.60 && -1.60 & \cr
& $B$ && -0.57 && -1.94 & \cr
& $D$ (Nijmegen) && -1.90 && -1.96 & \cr
& $F$ (Nijmegen) && -2.29 && -2.20 & \cr
& Expt. [19] && -15.0<a_s<0 && -3.2<a_t<-0.6 & \cr
& Expt. [20] && -4.3<a_s<1.0 && -2.0<a_t<-0.5 & \cr\noalign{\hrule}}}}$$

It should be noted that no attempt was made to ensure that the present
parameters give the best fit to the data which is possible in our models.
However, a very good fit is obtained with little effort.

In Figs.~8 and 9 we present the phase parameters for $\Lambda N\to\Lambda N$
scattering.  These are needed to predict all the spin observables, in addition
to the unpolarized differential and total cross sections presented here.  Such
spin observables may be experimentally determined in the near future.  We can
provide the phase parameters for this and all the other $\Lambda N-\Sigma N$
reactions as needed.  We also present these figures to describe the features
of channel coupling.  Of course, the elasticity parameter, $\eta$, decreases
from unity at the $\Sigma N$ threshold.  In addition, the phases, $\delta$,
and the mixing parameters, $\bar\epsilon$,
also show structure at this energy, particularly when
the partial wave couples to an $S$-state $\Sigma N$ channel (as do
the ${}^1S_0$ and
${}^3S_1-{}^3D_1$ $\Lambda N$ channels).  Similar effects occur at the
$\Sigma^*N$ and $\Sigma\Delta$ isobar channel thresholds,
substantially so except in the ${}^1S_0$, ${}^3P_0$ and ${}^1D_2-{}^3D_2$
$\Lambda N$ channels.

In addition to the threshold effects, the coupling adds attraction to the
scattering channel, increasing with energy up to the inelastic threshold.
This is of particular importance in the ${}^3S_1$ channel. As seen in Fig.~7
the $\Lambda N$ total elastic cross section becomes much too large at momenta
larger than 700~MeV/c in Model 1, without isobar coupling.  This can be traced
to the fact that $\delta\left( {}^3S_1\right)$ becomes very negative (Fig.~8)
reaching $-60^\circ$ at 1000~MeV/c.  We have added boundary condition coupling
to the $\Sigma^* N\left({}^3S_1\right)$ system, supplementing the meson
exchange potential coupling, so that the minimum value of
$\delta\left({}^3S_1\right)$ is $-24^\circ$.  This gives a good fit to the
higher-energy cross section while maintaining the low-energy fit.  When the
width of the isobars is taken into account the threshold effects will be less
sharp than shown here, but the attractive effect below threshold will be just
as strong.
\vfill
\eject
\bigskip
\noindent{\bf VI.\quad IMPLICATIONS FOR EXOTIC STRUCTURES}
\medskip
\nobreak
The bag Hamiltonian for the $\left[ q\left(1s_{1/2}\right)\right]^{N_n} \left[
s(1s_{1/2})\right]^{N_s}$ configurations confined to a radius $R$ is:$^{22}$
$$\eqalign{H &= {4\pi\over 3} BR^3 - {Z_0\over R} + N_n {\omega(m_qR)\over R}
+ N_s {\omega (m_sR)\over R}\cr
&+ {\alpha_s\over R} M_{ns} \left[ N (N-10) + 4\left( C_3 + {1\over 3} {\svec
J}^2\right)\right] \cr
&+ {\alpha_s\over R} \left( M_{nn} - M_{ns}\right) \left[ {3\over 4} N^2_n -
N_n - C_4 + 4\left( {\svec I}^2 + {1\over 3} {\svec J}^2_n\right)\right]\cr
&+ {\alpha_s\over R}\left( M_{ss} - M_{ns} \right) \left[ {3\over 2} N^2_s -
N_s - {2\over 3} {\svec J}^2_s\right] \cr
&+ {\alpha_s\over R}\left( E_{nn} - E_{ns}\right) \left[ {7\over 12} N_n
(12-N_n) - C_4\right] \cr
&+ {\alpha_s\over R} \left( E_{ss} - E_{ns}\right) \left[ {5\over 6} N_s
(6-N_s) - 2 {\svec J}^2_s\right]\ \ ,\cr}\eqno(6.1)$$
where the first term is the difference between perturbative and physical
vacuum energies determined by the ``bag constant'' energy density $B$.  The
second term is the cavity correction to the gluon self-energy and can also
absorb the center-of-mass motion correction (when that is not explicitly
calculated).  The third and fourth terms are the $1s_{1/2}$ Dirac state
energies of quarks of masses $m_q$ and $m_s$, respectively, where $N_n$ is the
number of non-strange quarks of current mass $m_q$, and $N_s$ is the number of
strange quarks of current mass $m_s(N_n + N_s = 3\times\hbox{number of
baryons})$.  The remaining terms arise from one-gluon exchange between quarks,
the first three being of color-magnetic type and the last two of
color-electric type.  In these terms $N = N_n + N_s$; $\alpha_s$ is the
strong coupling constant; the $M_{ij}$ and $E_{ij}$ are the
spatial matrix elements multiplied by $R$
for pairs of an $i$-quark and a $j$-quark (they depend
only on the products $m_iR$ and $m_jR$).  The ${\svec J}$ and ${\svec I}$ are
the total angular momentum and isospin of the configuration, while
${\svec J}_n$ and
${\svec J}_s$ refer to the total angular momentum of the non-strange and
strange
quarks, respectively.  $C_3$ is the quadratic Casimir operator in $SU(3,F)$
and $C_4$ in $SU(4,IJ_n)$.

Note that in the limit of equal mass quarks $M_{nn} = M_{ns} = M_{ss}$ and
$E_{nn} = E_{ns} = E_{ss}$.  Consequently, only the first of the gluon
exchange terms survives, and it depends only on the quantum numbers of the
total configuration.  The flavor mixing operator ${\svec J}^2_n$ occurs only
with the small coefficient ($M_{nn} - M_{ns}$), so that each state is
dominated by one flavor multiplet , the mass dependence on the mixing being
less than 14~MeV.  We ignore the flavor mixing, but present the resultant mass
uncertainty.  ${\svec J}^2_s$ also occurs with small difference coefficients
but is uniquely 3/4 for our case, $N_s=1$.

The mass spectrum of the $\Lambda N - \Sigma N$ exotic states, in
configurations of $1s_{1/2}$ quarks, is evaluated from Eq.~(6.1) using
$N_s=1$, $N_n = 5$.  The bag constants are those of the CBM, which were also
used for the nucleon-nucleon exotics.$^{4,\,5,\,6}$  Their values were
obtained by fitting the hadron spectrum$^{23}$ and are $B^{1/4} = 0.169$~MeV,
$Z_0 = 1.80$, $\alpha_s = 0.4225$ and $m_s = 0.181$~GeV.  We use $m_u = m_d =
0$.

The bag model boundary conditions determine the Dirac state eigenvalue
$\omega(m_qR)/R$, whose eigenfunctions result in closed expressions for the
$M_{ij}$ and $E_{ij}$.$^{24}$  The dependence of these quantities on $m_qR$ is
given in Figs.~2 and 3 of Ref.~[24] and is tabulated in Table~V of Ref.~[23].

The relevant Casimir operator values are $c_4(60) = {71/4}$, $c_3 (8) = 3$,
$c_3$(10 or 10$^*$) = 6, $c_3(27) = 8$ and $c_3(35)=12$.

With this input to Eq.~(6.1) we obtain the masses of the $B=2$, $S=-1$ states
with $(\dim F, J, I) = (8,1,1/2)$, $(8,2,1/2)$, $(10^*,1,1/2)$, $(10,1,3/2)$,
$(27,0,1/2)$, $(27,0,3/2)$, $(27,2,1/2)$, $(27,2,3/2)$, $(10^*,3,1/2)$, and
$(35,1,3/2)$, all of the $I=1/2$ and 3/2 states in the configuration.  They
were
each calculated for the three values of $R=3.04$, 5.575 and 8.11~GeV$^{-1}$.

A parabolic fit to these values locates the equilibrium (minimum) values of
the mass of each state.  The associated $R_{eq}$ increases from
5.79~GeV$^{-1}$ to 6.13~GeV$^{-1}$ as the equilibrium mass increases between
the $(8,1,1/2)$ and $(35,1,3/2)$ states.

Choosing the separation radius of the $R$-matrix method to be $R_0 =
5.21$~GeV$^{-1}$, the ratio $R_0/R_{eq} = 0.85-0.90$ for all the exotic
states.  On the other hand, the value $R_0 = 4.90$~GeV$^{-1}$ (corresponding
by Eq.~(3.1) to the value $r_0 = 1.082$~fm, used in $\Lambda N - \Sigma N$
scattering results of Section~V) results in $R_0/R_{eq} = 0.80-0.85$.  Both of
these choices of $R_0$ are consistent with the criteria for use of the
$R$-matrix method in this situation (Section~III).  When complete sets of
scattering data, enabling a phase shift analysis, become available at energies
approaching the exotic states, the value of $R_0$ can be more precisely
determined as it was in the $NN$ case.$^{4,\,5}$  In Table~VIII we present the
predictions of the exotic masses for both $R_0 = 4.90$~GeV$^{-1}$ and
5.21~GeV$^{-1}$.  Table IX lists the $\Lambda N$, $\Sigma N$,
$\Sigma(\Lambda)\Delta$, $\Sigma^*N$ and $\Sigma^*\Delta$ content (fractional
parentage coefficient squared) of each exotic state.$^{25}$
  In each case the sum of
the hadronic content is 0.2, just as in the $NN$ case of six non-strange
$s$-state quarks.$^7$  For each state Table~IX also lists $\rho_{\rm tot}$,
the sum of the residues to the hadronic states as given by Eq.~(3.4) when
$\xi^i_\alpha \xi^\beta_\alpha$ is replaced by 0.2.  This quantity determines
the width of the exotic resonance, for any given phase space.  In the case of
the ${}^1S_0$ $NN$ exotic,$^{4,\,5}$ $\rho_{\rm tot} = 0.268$~GeV results in
$\Gamma = 50$~MeV, while for the ${}^1D_2$ $NN$ exotic$^5$ $\rho_{\rm tot} =
0.312$ and $\Gamma=100$~MeV.  The approximate widths of the $S=-1$ exotics can
be inferred from the $NN$ results.  For the lowest mass $(8,1,1/2)$ state both
$\rho_{\rm tot}$ and the phase space are smaller than for the ${}^1S_0$ $NN$
case, so we may expect $\Gamma<50$~MeV.

\midinsert
\baselineskip 12pt plus 1pt minus 1pt
$$\hbox{\vbox{\offinterlineskip
\def\strut{\hbox{\vrule height 11pt depth 6pt width 0pt}}
\def\superstrut{\hbox{\vrule height 15pt depth 8pt width 0pt}}
\hrule
\halign{
\strut\vrule#&\tabskip .14in&
\hfil$#$\hfil&
\hfil$#$\hfil&
\hfil$#$\hfil&
\vrule#&
\hfil$#$\hfil&
\vrule#&
\hfil$#$\hfil&
\vrule#\tabskip 0.0in\cr
& \multispan7\hfil{\bf Table VIII:} Exotic Masses\hfil & \cr\noalign{\hrule}
& \multispan3\hfil State\hfil && \multispan3\hfil Masses of $B=2$, $S=-1$
Exotics in GeV\hfil & \cr\noalign{\hrule}
& \ \dim F &J & I && R_0 = 4.90\,\hbox{GeV}^{-1} && R_0 =
5.21\,\hbox{GeV}^{-1} & \cr\noalign{\hrule}
& \phantom{1}8 & 1 & 1/2 && 2.564 - 2.572 && 2.506 - 2.514 & \cr
& \phantom{1}8 & 2 & 1/2 && 2.644 - 2.657 && 2.580 - 2.593 & \cr
& 10 & 1 & 1/2 && 2.729 - 2.737 && 2.657 - 2.665 & \cr
& 10 & 1 & 3/2 && 2.753 - 2.765 && 2.681 - 2.693 & \cr
& 27 & 0 & 1/2 && 2.802 && 2.725 & \cr
& 27 & 0 & 3/2 && 2.826 && 2.748 & \cr
& 27 & 2 & 1/2 && 2.919 - 2.933 && 2.833 - 2.847 & \cr
& 10 & 3 & 1/2 && 2.933 && 2.847 & \cr
& 27 & 2 & 3/2 && 2.943 - 2.955 && 2.857 - 2.869 & \cr
& 35 & 1 & 3/2 && 3.083 - 3.095 && 2.984 - 2.997 & \cr\noalign{\hrule}}}}$$
\endinsert
\midinsert
\baselineskip 12pt plus 1pt minus 1pt
$$\hbox{\vbox{\offinterlineskip
\def\strut{\hbox{\vrule height 8.5pt depth 4.5pt width 0pt}}
\def\superstrut{\hbox{\vrule height 15pt depth 8pt width 0pt}}
\hrule
\halign{
\strut\vrule#&\tabskip .14in&
\hfil$#$\hfil&
\hfil$#$\hfil&
\hfil$#$\hfil&
\vrule#&
\hfil$#$\hfil&
\hfil$#$\hfil&
\hfil$#$\hfil&
\hfil$#$\hfil&
\vrule#&
\hfil$#$\hfil&
\vrule#&
\hfil$#$\hfil&
\vrule#\tabskip 0.0in\cr
&\multispan{12}\hfil{\bf Table IX:} Hadronic Content\hfil & \cr\noalign{\hrule}
\superstrut& \multispan3\hfil State\hfil && \multispan4\hfil
$\matrix{\hbox{Square of Fractional}\cr\hbox{Percentage Coefficient,}\
\xi^2\cr}$
\hfil && \multispan3\hfil $\rho_{\rm tot}$~(GeV)\hfil &
\cr\noalign{\hrule}
\superstrut
& \ \dim F & J & I && \Sigma N(\Lambda N) & \Sigma\Delta(\Lambda\Delta) &
\Sigma^*N & \Sigma^*\Delta && \matrix{R_0 = 4.90\cr\hbox{GeV}^{-1}\cr}
 && \matrix{R_0 =
5.21\cr\hbox{GeV}\cr} & \cr\noalign{\hrule}
& \phantom{1}8\phantom{^*}
& 1 & 1/2 && 4/45 & 1/18 & 1/18 & && 0.225 && 0.156 & \cr
& \phantom{1}8\phantom{^*} & 2 & 1/2 && & 1/10 & 1/10 & && 0.242 && 0.173 & \cr
& 10^* & 1 & 1/2 && 1/9\phantom{1} & & & 4/45 && 0.268 && 0.198 & \cr
& 10\phantom{^*} & 1 & 3/2 && 1/45 & 4/45 & 4/45 & && 0.268 && 0.198 & \cr
& 27\phantom{^*} & 0 & 1/2 && 1/9\phantom{1} & & & 4/45 && 0.288 && 0.217 & \cr
& 27\phantom{^*} & 0 & 3/2 && 1/9\phantom{1} & & & 4/45 && 0.288 && 0.217 & \cr
& 27\phantom{^*} & 2 & 1/2 && & 2/45 & 2/45 & 1/9\phantom{1} && 0.315 && 0.244
&
 \cr
& 10^* & 3 & 1/2 && & & & 1/5\phantom{1} && 0.315 && 0.244 & \cr
& 27\phantom{^*} & 2 & 3/2 && & 2/45 & 2/45 & 1/9\phantom{1} && 0.316 && 0.245
&
 \cr
& 35\phantom{^*} & 1 & 3/2 && & 2/45 & 2/45 & 1/9\phantom{1} && 0.356 && 0.283
&
\cr\noalign{\hrule}}}}$$
\endinsert

In Fig.~10 the distribution of exotic masses (for $R_0 =
4.90\,\hbox{GeV}^{-1}$) is shown relative to the two-baryon thresholds.  The
lowest mass exotic at 2.506~GeV (for $R_0 = 5.21\,\hbox{GeV}^{-1})$
is above all thresholds except that of the
$\Sigma^*\Delta$ (2.620~GeV), and only the two $\dim F=8$ exotics may be below
this threshold.  The width of this threshold structure is likely to be
$>200$~MeV, so even for these lightest exotics it should be possible to
disentangle exotic and threshold structures.

These results indicate the likely presence of a rich structure of $S=-1$
exotics detectable with $\Lambda$ and $\Sigma$ beams with laboratory energy in
the 1--3~GeV range.  In future work we will include the $f$-poles due to the
exotic states in the scattering analysis.  This will provide prediction of the
widths and inelasticities of the exotic structures, and will show the
amplitudes of these structures for the different spin observables.
\goodbreak
\bigskip
\centerline{\bf ACKNOWLEDGEMENTS}
\medskip
One of us, ELL, benefited from the facilities and many conversations while
visiting the Physics Division of Los Alamos National Laboratory, and the
Institute f\"ur Kernphysik at KFA/J\"ulich.  Discussions with Professor
J.~Speth, Dr. K. ~Holinde, Dr.~G. Stephenson and Dr.~T.~Goldman were
informative and stimulating.  In particular, Dr.~B.~Holzenkamp provided
important information during and after his thesis work on the momentum space
meson-exchange model of the hyperon-nucleon interaction.
\vfill
\eject
\centerline{\bf REFERENCES}
\medskip
\item{1.}B. Holzenkamp, K. Holinde and J. Speth, {\it Nucl. Phys.\/} {\bf
A500}, 485 (1989).
\medskip
\item{2.}E. L. Lomon, {\it Phys. Rev.\/} {\bf D26}, 576 (1982).
\medskip
\item{3.}P. Gonz\'alez and E. L. Lomon, {\it Phys. Rev.\/} {\bf D34}, 1351
(1986).
\medskip
\item{4.}P. LaFrance and E. L. Lomon, {\it Phys. Rev.\/} {\bf D34}, 1341
(1986).
\medskip
\item{5.}P. Gonz\'alez, P. LaFrance and E. L. Lomon, {\it Phys. Rev.\/} {\bf
D35}, 2142 (1987).
\medskip
\item{6.}E. L. Lomon, {\it Colloque de Physique\/} {\bf 51}, suppl\'ement 22,
15 November 1990, C6-363, published by Les Editions de Physique.
\medskip
\item{7.}E. L. Lomon, in {\it AIP Conference Proceedings\/} {\bf 187}, 630
(1989).
\medskip
\item{8.}M. Creutz, {\it Phys. Rev. Lett.\/} {\bf 45}, 313 (1980).
\medskip
\item{9.}M. M. Nagels, T. A. Rijken, J. J. deSwart, {\it Phys. Rev.\/} {\bf
D12}, 744 (1975); {\it Phys. Rev.\/} {\bf D15},2547 (1977);
{\it Phys. Rev.\/} {\bf D20}, 1633 (1979).
\medskip
\item{10.}W. Grein and P. Kroll, {\it Nucl. Phys.\/} {\bf A338}, 332 (1980).
\medskip
\item{11.}K. Holinde, private communication.
\medskip
\item{12.}E. P. Wigner and L. Eisenbud, {\it Phys. Rev.\/} {\bf 72}, 29 (1947);
A. M. Lane and R. A. Thomas, {\it Rev. Mod. Phys.\/} {\bf 30}, 257 (1958).
\medskip
\item{13.}R. J. Jaffe and F. E. Low, {\it Phys. Rev.\/} {\bf D19}, 2105
(1979).
\medskip
\item{14.}I. P. Auer {\it et al.\/}, {\it Phys. Rev. Lett\/} {\bf 62}, 2649
(1989); R. Bertini {\it et al,\/} {\it Phys. Lett.\/} {\bf 203}, 18 (1988).
\medskip
\item{15.}M. L. Goldberger, M. T. Grisaru, S. W. MacDowell and D. Y. Wong,
{\it Phys. Rev.\/} {\bf 120}, 2250 (1960); J. Bystricky, F. Lehar and P.
Winternitz, {\it J. Physique\/} {\bf 39}, 1 (1978); J. Bystricky, C.
Lechanoine-Leluc and F. Lehar, {\it J. Physique\/} {\bf 48}, 199 (1987).
\medskip
\item{16.}F. Eisele, H. Filthuth, W. Fohlisch, V. Hepp, E. Leitner and G. Zech,
{\it Phys. Lett.\/} {\bf 37B}, 204 (1971).
\medskip
\item{17.}R. Engelmann, H. Filthuth, V. Hepp and E. Kluge, {\it
Phys. Lett.\/} {\bf 21}, 587 (1966).
\medskip
\item{18.}J. J. deSwart and C. Dullemond,  {\it Ann. Phys.\/} {\bf 19}, 458
 (1962).
\medskip
\item{19.}V. Hepp and H. Schleich,  {\it Z. Phys.\/} {\bf 214}, 71 (1968).
\medskip
\item{20.}G. Alexander {\it et al.\/},
{\it Phys. Rev.\/} {\bf 173}, 1452 (1968); also {\it Phys. Rev. Lett.\/} {\bf
7}, 348 (1961); J. A. Kadyk {\it et al.\/}, {\it Nucl. Phys.\/} {\bf B27}, 13
(1971).
\medskip
\item{21.}B. Sechi-Zorn, B. Kehoe, J. Twitty and R. Burnstein, {\it
Phys. Rev.\/} {\bf 175}, 1735 (1968).
\medskip
\item{22.}A. Th. M. Aerts, P. J. G. Mulders and J. J. deSwart, {\it Phys.
Rev.\/} {\bf D17}, 260 (1978).
\medskip
\item{23.}P. J. Mulders and A. W. Thomas, {\it J. Phys.\/} {\bf G9}, 1159
(1983).
\medskip
\item{24.}T. DeGrand {\it et al.\/}, {\it Phys. Rev.\/} {\bf D12}, 2060
(1975).
\medskip
\item{25.}J. J. deSwart, private communication; M Harvey, {\it Nucl. Phys.\/}
{\bf A352}, 301 (1981), table 11; (E) {\it Nucl. Phys.\/} {\bf A481}, 834
(1988).
\medskip
\item{26.}A. deShalit and I. Talmi, {\it Nuclear Shell Theory\/} (Academic
Press, 1963).
\vfill
\eject
\centerline{\bf APPENDIX A}
\medskip
\centerline{\bf Isospin, Spin, and Tensor Matrix Elements}
\bigskip
\nobreak
The meson exchange potentials are proportional to the second order
perturbation amplitudes represented by the Feynman diagrams of Fig.~1
and 2. The evaluation of the diagrams and their Fourier transformation
results in the dependence of the relative coordinate $r$ given in
Eqs.~2.4, 2.9 and 2.10. It also results in matrix elements of the scalar
products of the vertex operators on baryon lines 1 and 2,
$$
\Bigl\langle I_1 I_2 I M_I S_1 S_2 L J M_J \Bigl| T^{(0)}_{12} \Bigr|
I'_1 I'_2 I M_I S'_1 S'_2 L' J M_J \Bigr\rangle \ \ ,
\eqno(\rm A.1)
$$
in which $T^{(0)}_{12}$ is one of four types of operators:
$$
T^{(0)}_{12} = \left[ T^k_1 \times T^k_2 \right]^{(0)} \ \ , \eqno(A.2a)
$$
where $T^k_i$ is a spin or isospin tensor of rank $k$ acting at vertex
$i$, including  transition (iso)spins,
$$
T^{(0)}_{12} =
S^T_{12} \equiv 3 (\svec S_1 \cdot \hat r) (\svec S_2 \cdot \hat r)
- \svec S_1 \cdot \svec S_2 \ \ , \eqno(A.2b)
$$
where the $\svec S_i$ are spin or transition spin operators and $\svec
r = \svec r_2 - \svec r_1$,
$$
\eqalignno{
T^{(0)}_{12} &= ( \svec \sigma_1 \pm \svec \sigma_2 ) \cdot \svec L \ \ ,
&(A.2c)\cr
\noalign{\hbox{or}}
T^{(0)}_{12} &= i ( \svec \sigma_1 \times \svec \sigma_2) \cdot
\svec L \ \ , &(A.2d)\cr}
$$
with the usual Pauli spin, $\svec \sigma$, and orbital angular
momentum, $\svec L$, operators.

The matrix elements of Eq.\ A.2c with the plus sign are readily evaluated using
${\svec J}^2 = {\svec L}^2 + {\svec S}^2 + 2 \svec S \cdot \svec L$,
$S = {1 \over 2} ( \svec \sigma_1 + \svec \sigma_2 )$ giving,
$$
\left\langle S L J \left| 2 \svec S \cdot \svec L \right| S'
L' J \right\rangle = S \left[ J (J+1) - L (L+1) - 2 \right]
\delta_{LL'} \delta_{SS'} \ \ . \eqno(A.3)
$$
Equation A.2c with the minus sign and Eq.~A.2d are efficiently
evaluated using the properties of $\sigma_{+, -, z}$ and
$L_{+, -, z}$ giving,
$$
\eqalignno{
\left\langle S L J \bigl|(\svec \sigma_1 - \svec \sigma_2) \cdot \svec L
\bigr| S' L' J \right\rangle &= -2 \sqrt{J(J+1)}
\bigl[1 - \delta_{SS'} \bigr] \delta_{LL'} \ \ , &(A.4)\cr
\noalign{\hbox{and}}
\left\langle S L J \bigl| i (\svec \sigma_1 \times \svec \sigma_2) \cdot
\svec L
\bigr| S' L' J \right\rangle &= 4 \sqrt{J(J+1)}
\bigl[1 - \delta_{SS'} \bigr] \left[\delta_{S'0}-{1 \over 2}
\right] \delta_{LL'} \ \ . &(A.5)
\cr}
$$
Diagonal (non-transition) spin or isospin operators have rank 1 and
A.2a can be expressed in terms of the scalar product$^{26}$
$\svec T_1 \cdot \svec T_2 \equiv - \sqrt3 \left[ T^{(1)}_1 \cdot
T^{(1)}_2 \right]^0$, enabling those matrix elements to be evaluated
from $2 \svec T_1 \cdot \svec T_2 = \left| \svec T_1 + \svec T_2 \right|^2
- \svec T^2_1 - \svec T^2_2$, just as for Eq.~A.3. This is not
possible for transition spin and the general case of A.2a is evaluated
using the relation for matrix elements of tensor products of operators
acting on different coordinates$^{26}$:
$$
\eqalignno{
\left( \alpha_1 j_1 \alpha_2 j_2 J \bigm| \bigm| T^{(0)}_{12} \bigm| \bigm|
\alpha_1' j_1' \alpha_2' j_2' J' \right)&=\,(-1)^{j_2 + J + j_1' + k}
\delta_{JJ'}\sqrt{2J+1 \over 2k+1}\cr
&\qquad \times \left\{ \matrix{j_1&j_2&J\cr
j_2'&j_1'&k\cr}\right\}
\left(\alpha_1 j_1 \bigl| \bigl| T^{(k)}_1 \bigr|\bigr|\alpha_1' j_1'\right)\cr
&\qquad \times
\left(\alpha_2 j_2 \bigl| \bigl| T^{(k)}_2 \bigr|\bigr|\alpha_2' j_2'\right)
 \ \ , &(A.6)\cr}
$$
in which the reduced matrix elements are defined by the Wigner-Eckart
theorem,
$$
\left\langle JM \bigl| T^{(k)}_\kappa \bigr| J'M' \right\rangle = \,(-1)^{J-M}
\left( \matrix{J&k&J'\cr -M&\kappa&M'\cr} \right)
\left( J \bigl|\bigl| T^{(k)} \bigr|\bigr| J' \right) \ \ , \eqno(A.7)
$$
and the $\left\{ \matrix{ j_1&j_2&J\cr j_2'&j_1'&k\cr} \right\}$ and
$\left( \matrix{ J&k&J'\cr -M&\kappa&M'\cr} \right)$ are Racah $(6j)$ and
Wigner $(3j)$ coefficients respectively. The $j_i (j_i')$ stand for
either the spin or isospin of the baryon entering (leaving) the vertex
on line $i$. The index $k$ is the angular momentum (or isospin)
carried into the vertex by the exchange meson. For the non-trivial
($k=0$) cases this is $k=1$ when the $\Delta j_i = 0$ or $1$ and
$k = {1 \over 2}$ when $\Delta j_i = {1 \over 2}$. The matrix elements
of Eq.~A.2a then reduce to the evaluation of the reduced matrix
elements of the (iso)spin or transition (iso)spin operators, which we
shall discuss at the end of this appendix.

In the operator of Eq.~A.2b, the scalar product is well defined
because the relevant spin and transition spin operators are all of
rank 1. This operator depends both on the independent spin coordinate
at vertex 1 and vertex 2, but also on the coordinate $\svec r$ which
depends on both vertices. We must first separate these two type of
dependence to apply the theorem relating matrix elements of the whole
tensor to matrix elements of its factors. Using the rearrangement
formula of p.\ 133 of Ref.~[26] we have,
$$
\eqalignno{
3 ( \svec S_1 \cdot \hat r )( \svec S_2 \cdot \hat r )
&\equiv 9 [ \svec S_1 \times \hat r ]^0
 \left[ \svec S_2 \times \hat r \right]^0 \ \ , \cr
&= 9 \sum_\lambda (-1)^\lambda \sqrt{2\lambda+1}
 \left\{ \matrix{ 1&1&\lambda\cr 1&1&0\cr}\right\}
 \left[ [ \svec S_1 \times \svec S_2 ]^\lambda \times
 [ \hat r \times \hat r ]^\lambda \right]^0 \ \ , \cr
&= 3 \sum_{\lambda = 0,2} \sqrt{2\lambda+1}
 \left[ [ \svec S_1 \times \svec S_2 ]^\lambda \times
 [ \hat r \times \hat r ]^\lambda \right]^0 \ \ , \cr
&= \svec S_1 \cdot \svec S_2 + 3\sqrt5
 \left[ [ \svec S_1 \times \svec S_2 ]^2
 [ \hat r \times \hat r ]^2 \right]^0 \ \ , &(A.8)\cr}
$$
where we have used $\hat r \cdot \hat r = 1$ and
$[\hat r \times \hat r]^1 = 0$.

It follows that the operator of Eq.~A.2b can be expressed as a
tensor product of
operators of different coordinates,
$$
3(\svec S_1\cdot\hat r)(\svec S_2\cdot\hat r) - \svec S_1\cdot\svec S_2
= 3\sqrt5 \left[[\svec S_1 \times\svec S_2]^2 [\hat r \times \hat r]^2
\right]^0 \ \ . \eqno(A.9)
$$
 Eq.~A.6 may be applied
to reduce the matrix element of Eq.~A.2b to matrix elements of
${[\svec S_1 \times \svec S_2]^2}$ and ${[\hat r \times \hat r]}$. To the
first of these one can apply Eq.~A.6 again, while the second requires the
expression for matrix elements of a tensor product depending on the
same coordinate$^{26}$,
$$
\eqalignno{
\left( \alpha j \bigl| \bigl| T^k \bigr| \bigr| \alpha' j' \right)
&= (-1)^{j+k+j'} \sqrt{2k+1} \sum_{\alpha'' j''}
 \left( \alpha j \bigl|\bigl| T^{(k_1)} \bigr|\bigr|\alpha'' j'' \right)
 \left( \alpha'' j'' \bigl|\bigl|T^{(k_2)}\bigr|\bigr|\alpha' j' \right)\cr
&\qquad \times \left\{ \matrix{k_1&k_2&k\cr j'&j&j''\cr}\right\}
\ \ . &(A.10)\cr}
$$

After performing the sums one obtains,
$$
\eqalignno{
\left\langle S_1 S_2 S L J \bigl| S^T_{12}\bigr| S_1' S_2' S' L' J\right\rangle
&= (-1)^{S+J} \bigl[ 30 (2L+1)(2L'+1)(2S+1)(2S'+1)\bigr]^{1\over 2}\cr
&\qquad \times \left\{\matrix{J&S&L\cr 2&L'&S'\cr}\right\}
 \left(\matrix{L&2&L'\cr 0&0&0\cr}\right)
 \left\{\matrix{S_1&S_1'&1\cr S_2&S_2'&1\cr S&S'&2\cr}\right\}\cr
&\qquad \times \left( S_1 \bigl|\bigl| \svec S_1 \bigr|\bigr| S_1' \right)
 \left( S_2 \bigl|\bigl| \svec S_2 \bigr|\bigr| S_2' \right)
\ \ . &(A.11)\cr}
$$

With the above relations all of the required matrix elements are given
once the reduced matrix elements of the ordinary and transition spin
and isospin operators are known. These are obtained from the
definitions of these operators and the Wigner-Eckart theorem, Eq.~A.7. The
tensor properties of all the spin operators requires proportionality to the
Clebsch-Gordon coefficients
${\langle J''\, M'' | J \, M \, J'\, M' \rangle}$,
$$
\left\langle S_2 \, M_2 | S^{M_j} | S_1\, M_1 \right\rangle
= N \left\langle S_2\, M_2 \bigm| j\, M_j\, S_1\, M_1
\right\rangle \ \ , \eqno(A.12)
$$
where $j=1 \left({1\over 2}\right)$ when $| S_1 - S_2 |$ is integer
(half-odd-integer).  For all the transition spin operators $N=1$, while
for ordinary spin operators $N$ is determined by
$\svec S \cdot \svec S = S(S+1)$. For example,
$\svec \sigma = 2 {\svec S}_{{1\over 2}, {1\over 2}}$ requires,
$$
\left\langle {1\over 2} \, M' | \sigma^{M_j} | {1\over 2} \, M
\right\rangle = -2 \sqrt{3\over 4}
\left\langle {1\over 2}\, M' \bigm| 1\, M_j\, {1\over 2}\, M \right\rangle
. \eqno(A.13)
$$

The following results satisfy our needs:
$$
\displaylines{
\qquad\left( {1\over 2} \Bigm|\Bigm| \svec\sigma\Bigm|\Bigm| {1\over 2} \right)
  = \sqrt6\kern 50pt
\left( {3\over 2}\Bigm|\Bigm| \svec \Sigma \Bigm|\Bigm| {3\over 2}\right)
  = 2 \sqrt{15} \qquad \hbox{where} \qquad \svec \Sigma=2 {\svec S}_{{3\over
2},
{3\over2}}\hfill\cr
\qquad\left( 1 \Bigm|\Bigm| \svec
S_{1,1} \Bigm|\Bigm| 1 \right) = \sqrt6\hfill\cr
\qquad \left( {3\over 2}\Bigm|\Bigm| \svec S_{{1\over 2},{3\over 2}}
  \Bigm|\Bigm| {1\over 2} \right)
  = -\left( {1\over 2} \Bigm|\Bigm| \svec S_{{3\over 2},{1\over 2}}
  \Bigm|\Bigm| {3 \over 2}\right) = 2\hfill\cr
\qquad \left( 1 \Bigm|\Bigm| \svec S_{0,1}\Bigm|\Bigm| 0 \right) = \sqrt3
\kern 50pt
\left( 0 \Bigm|\Bigm| \svec S_{1,0}\Bigm|\Bigm| 1 \right)= 1\hfill\cr
\qquad \left( 1 \Bigm|\Bigm| \svec
S_{{1\over 2},1}\Bigm|\Bigm| {1\over 2}\right)
  = -\sqrt3\kern 50pt
\left( {3\over 2}\Bigm|\Bigm|\svec S_{1, {3\over 2}}\Bigm|\Bigm| 1 \right)
  = -2\hfill\cr
\qquad \left( {1\over 2}\Bigm|\Bigm| \svec
I_{0,{1\over 2}}\Bigm|\Bigm| 0 \right)
  = -\sqrt2\kern 50pt
\left( 0 \Bigm|\Bigm| \svec I_{{1\over 2}, 0}\Bigm|\Bigm| {1\over 2}\right)
  = -1\hfill\cr}
$$
\vfill
\eject
\centerline{\bf FIGURE CAPTIONS}
\medskip

\item{Fig.~1:} Feynman diagrams generating the potential matrix:
(a)~${\Lambda N\rightarrow} {\Lambda N}$;
(b)~${\Lambda N\rightarrow} {\Sigma N}$ (and time
reversal of ${\Sigma N\rightarrow} {\Lambda N}$); (c)~${\Sigma N\rightarrow}
{\Sigma N}$; (d)~${\Lambda N\rightarrow} {\Sigma\Delta}$;
(e)~${\Lambda N\rightarrow}{\Sigma^*N}$; (f)~${\Sigma N\rightarrow}
{\Lambda\Delta}$; (g)~${\Sigma N\rightarrow}
{\Sigma\Delta}$; (h)~${\Sigma N\rightarrow} {\Sigma^*N}$.
Processes (d--h) become part of the ${\Lambda N\rightarrow}
{\Sigma N}$ scattering studied here when iterated by the Schr\"odinger
equation with the time reversed diagrams.
\medskip
\item{Fig.~2:} The basic Feynman diagram showing the line labelling and
kinematics in the center-of-momentum frame.
\medskip
\item{Fig.~3:} The meson exchange potentials which contribute to the
Feynman diagrams in Fig.~1(c).
The central potentials ($V_c$) are shown in
solid, the spin-orbit ($V_{SL}$) are shown as the dashed curves, the spin-spin
($V_\sigma$) potentials are dotted, the tensor ($V_T$) potentials are shown as
the dot-dashed curves, and, finally, the antisymmetric spin-orbit potentials
($V_{SLA}$ and $V_{SSL}$) are
displayed as the closely spaced dotted curves and the  closely
spaced, long dashed curves, respectively.
\medskip
\item{Fig.~4:} The meson exchange potentials which contribute to the
${\Sigma N\rightarrow} {\Lambda\Delta}$ interaction, Fig.~1(f).
The notation is as in Fig.~3.
\medskip
\item{Fig.~5:} ${\Sigma^+p\rightarrow} {\Sigma^+p}$: comparison of models with
experiment.  The data is from Ref.~[16].  The dotted, dashed, and solid
curves correspond to Models 1 (no isobar coupling, no coulomb), 2 (isobar
coupling, no coulomb), and 3 (isobar coupling, coulomb) respectively.
(a)~Total cross section.  (b)~Differential cross section at
${p_{\Sigma^+}=}170~MeV/c$.
\medskip
\item{Fig.~6:} ${\Sigma^-p}$ induced reactions: comparison of models with
experiment.  The curves are related to the models as in Fig.~5.
(a)~${\Sigma^-p\rightarrow} {\Sigma^-p}$ total cross section.  The data is from
Refs.~[16] and [17].  (b)~${\Sigma^-p\rightarrow} {\Sigma^-p}$ differential
cross section at ${p_{\Sigma^-}=}160~MeV/c$.  The data is from Ref.~[16].
(c)~${\Sigma^-p\rightarrow} {\Sigma^0n}$ total cross section.  The data is from
Refs.~[16] and [17].  (d)~${\Sigma^-p\rightarrow} {\Sigma^0n}$ differential
cross section at ${p_{\Sigma^-}=}160~MeV/c$. There is as yet no data.
(e)~${\Sigma^-p\rightarrow} {\Lambda n}$ total cross section.  The data is from
Ref.~[17].  (f)~${\Sigma^-p\rightarrow} {\Lambda n}$ differential cross section
at ${p_{\Sigma^-}=}160~MeV/c$.  The data is from Ref.~[17].
\medskip
\item{Fig.~7:} ${\Lambda p}$ induced reactions: comparison of models with
experiment.  The curves are related to the models as is Fig.~15.
(a)~The ${\Lambda p\rightarrow} {\Lambda p}$ total cross section.  The momentum
range is extended beyond the $\Sigma$ production threshold.  The data is from
Refs.~[20] and [21].  (b)~The ${\Lambda p\rightarrow} {\Sigma^0}p$ total cross
section.  The data is from Ref.~[20].
\medskip
\item{Fig.~8:} $I=1/2$ phase parameters $\delta$ and $\bar \epsilon$
 of the models for the
${\Lambda p\rightarrow} {\Lambda p}$ or ${\Sigma^0p}$ reactions.
The momentum range of 1750~$MeV/c$ is
large enough to include the ${\Sigma N}$, ${\Sigma^*N}$ and ${\Sigma\Delta}$
 thresholds, as indicated by arrows.  The curves denote the models as in
Fig.~5.  The phase parameters graphed are defined in the text.
\medskip
\item{Fig.~9:} $I=1/2$ phase parameters $\eta$ of the models
for the ${\Lambda p \rightarrow} {\Lambda p}$ or ${\Sigma^0 p}$ reactions.
The comments of Fig.~8 apply.
\medskip
\item{Fig.~10:}The mass distribution of the strangeness $=-1$, $B=2$ exotic
structures and the two-baryon thresholds. $---$ elastic thresholds; $\cdots$
isobar channel thresholds; --- exotic masses.  The double bars indicate the
uncertainty in exotic masses due to flavor mixing.  For $I=1/2$ the $(10^*,3)$,
which has no flavor mixing, is degenerate with the top of the $(27,2)$ range.
\par
\vfill
\end